\documentclass{article}

\usepackage{verbatim} 
\usepackage{float}

\usepackage{PRIMEarxiv}

\usepackage[utf8]{inputenc} 
\usepackage[T1]{fontenc}    
\usepackage{hyperref}       
\usepackage{url}            
\usepackage{booktabs}       
\usepackage{amsfonts}       
\usepackage{nicefrac}       
\usepackage{microtype}      
\usepackage{lipsum}
\usepackage{fancyhdr}       
\usepackage{graphicx}       

\usepackage{mathtools}
\usepackage{amsthm,amsmath,amssymb,amsfonts,exscale,latexsym,float,eucal}

\usepackage{lineno}
\usepackage{soul}
\usepackage{color, xcolor}
\usepackage{mathrsfs,amsmath}
\usepackage[linesnumbered,ruled]{algorithm2e}
\usepackage{cite}
\usepackage{hyperref}
\usepackage{cleveref}
\hypersetup{colorlinks = true,
	citecolor = blue}
	
\usepackage{upgreek}
\usepackage[utf8]{inputenc}
\usepackage[T1]{fontenc}

\title{Differentiable Imaging Meets Adaptive Neural Dropout: An Advancing Method for Transparent Object Tomography}

\author{
  Delong Yang \\
  School of Optics and Photonics, Beijing Institute of Technology \\
  Beijing, China \\
   \And
  Shaohui Zhang*\\
  School of Optics and Photonics, Beijing Institute of Technology \\
  Beijing, China \\
  \texttt{*zhangshaohui@bit.edu.cn} \\
   \And
  Jiasong Sun \\
  School of Electronic and Optical Engineering, Nanjing University of Science and Technology \\
  Nanjing, China \\
   \And
  Chao Zuo \\
  School of Electronic and Optical Engineering, Nanjing University of Science and Technology \\
  Nanjing, China \\
   \And
  Qun Hao*\\
  School of Optoelectronic Engineering, Changchun University of Science and Technology \\
  Changchun, China
  \texttt{*qhao@bit.edu.cn}
}

\begin{document}
\maketitle

\begin{abstract} 
Label-free tomographic microscopy offers a promising alternative for visualizing biological structures, enabling the reconstruction of a three-dimensional (3D) refractive index (RI) distribution from two-dimensional (2D) intensity measurements. However, the accuracy of the forward model and the ill-posed nature of the inverse reconstruction problem pose significant challenges to obtaining artifact-free RI maps. In recent years, artificial neural networks have garnered significant attention for their powerful nonlinear fitting capabilities.
In this work, we incorporate a Differentiable Imaging approach, representing the 3D sample as a multi-layer neural network that embeds the physical constraints of light propagation at every layer. Building upon this integrated approach, we propose a physics-guided Adaptive Dropout Neural Network (ADNN) optimization method for optical diffraction tomography (ODT). Rather than focusing on conventional input–output mapping, our framework emphasizes network topology and voxel-wise complex RI fidelity. By leveraging prior knowledge of the sample’s refractive index, the ADNN dynamically drops out certain neurons during each optimization iteration and selectively reactivates previously dropped neurons in subsequent iterations, thereby enhancing the accuracy and stability of the reconstruction at a granular level.
We validate the proposed method through extensive simulations and experiments on various sample types (spanning weakly scattering and multiple scattering regimes) and different imaging setups. The results demonstrate excellent performance and broad applicability for advancing label-free 3D biological imaging. This unique strategy not only enhances the quantitative accuracy of 3D RI reconstructions but also provides superior optical-sectioning capabilities, effectively suppressing artifacts. Specifically, experimental results show that the ADNN reduces the Mean Absolute Error (MAE) by a factor of 3 to 5 and improves the Structural Similarity Index Metric (SSIM) by approximately 4 to 30 times compared to the state-of-the-art method.
\end{abstract}

In biological imaging, acquiring three-dimensional (3D) structural information from transparent or semi-transparent samples is of critical importance. Such 3D information offers powerful insights into various fields, including morphogenesis\textsuperscript{\cite{keller2013imaging}}, oncology\textsuperscript{\cite{valls2023exploring}}, cellular pathophysiology\textsuperscript{\cite{kim2018measurements}}, and biochemistry\textsuperscript{\cite{yla20093d}}, allowing researchers to investigate complex biological processes and deepen our understanding of disease mechanisms and cellular functions.

When combined with 3D mechanical scanning, fluorescence microscopy techniques\textsuperscript{\cite{agard1989fluorescence,lichtman2005fluorescence,huang2009super}}—such as STED\textsuperscript{\cite{vicidomini2018sted}}, PALM\textsuperscript{\cite{shroff2008live}}, and STORM\textsuperscript{\cite{huang2008three}}—enable 3D imaging of transparent or semi-transparent biosamples by detecting fluorescent markers. However, these approaches require external labeling, which can lead to phototoxicity and photobleaching\textsuperscript{\cite{hoebe2008quantitative}}, and not all structures are effectively labeled\textsuperscript{\cite{perfilov2021transient,pawley200039}}.

Phase imaging provides a label-free alternative, detecting phase variations that reveal structural information in transparent or semi-transparent samples\textsuperscript{\cite{park2018quantitative}}. These techniques can be categorized into interference-based and non-interference-based methods\textsuperscript{\cite{mir2012quantitative}}. Interference-based methods (e.g., classical digital holographic microscopy\textsuperscript{\cite{mico2008common,el2018quantitative}}) offer high-precision phase retrieval but tend to be costly\textsuperscript{\cite{picazo2022design}}, complex, and sensitive to environmental fluctuations\textsuperscript{\cite{zhang2021review}}. Non-interference approaches depend solely on intensity measurements\textsuperscript{\cite{zheng2013wide,jingshan2014transport}}, making them more robust and cost-effective, but phase variations accumulate along the axial direction\textsuperscript{\cite{nguyen2022quantitative}}, rendering direct 3D imaging difficult.

To circumvent this limitation, researchers have shown that multi-angle illumination intensity stacks in non-interference systems implicitly carry 3D structural information\textsuperscript{\cite{horstmeyer2016diffraction}}. Linking the sample’s 3D refractive index (RI) with the 2D light field on the sensor can thus enable 3D tomographic imaging. However, the complexity of light–sample interactions\textsuperscript{\cite{taflove2005computational}} and the limited redundancy in multi-angle illumination data\textsuperscript{\cite{zhou2022transport}} lead to a highly nonlinear and severely ill-posed inverse problem\textsuperscript{\cite{arridge2009optical}}.

\begin{figure}[h]
	\centering
	\includegraphics[scale=0.32]{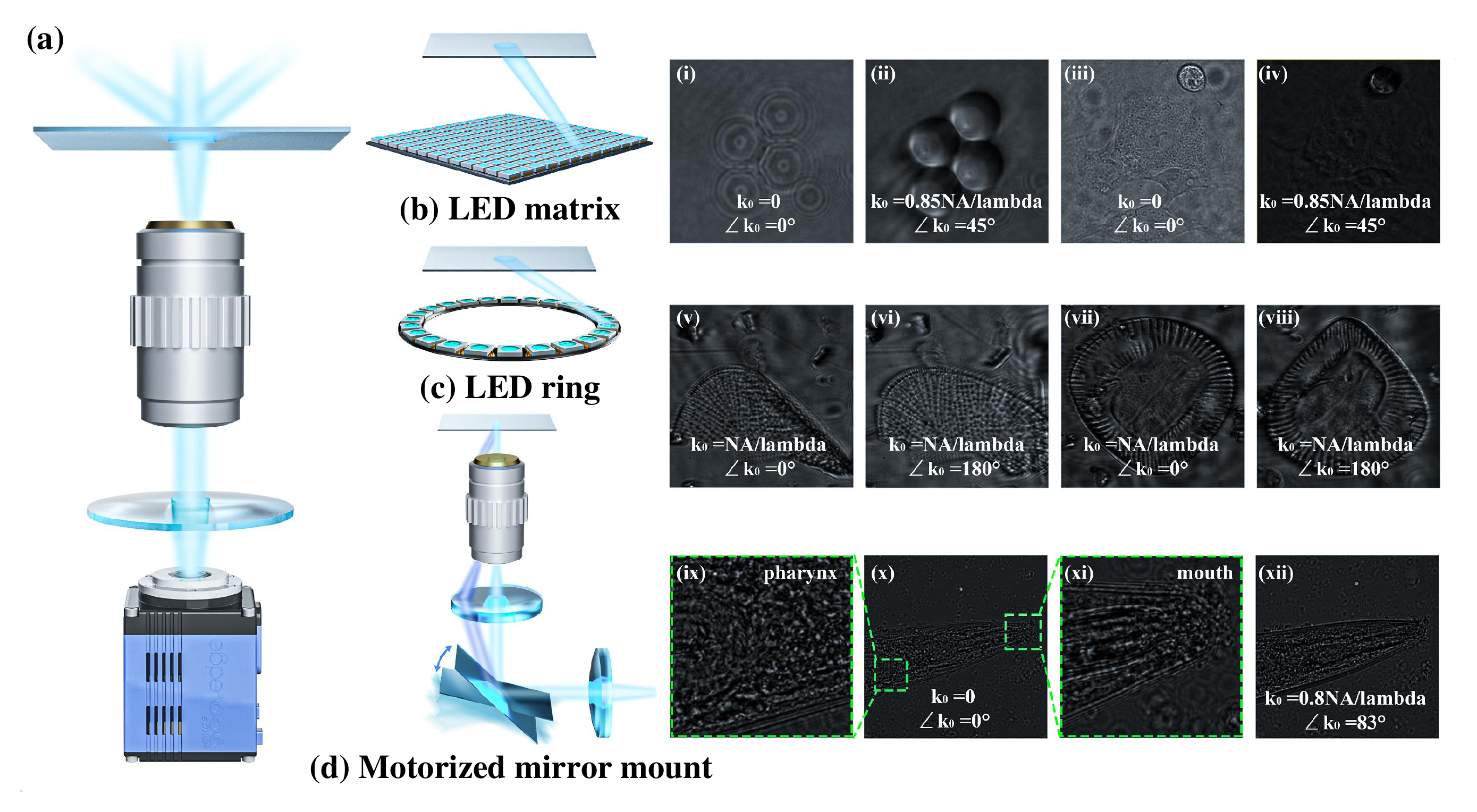}
	\caption{(a) Non-interference microscopy imaging system, with multi-angle illumination provided by different programmable light sources. (b) LED matrix and the corresponding raw image, where multi-angle illumination is provided by sequential activation from the center in a spiral pattern. (c) LED ring and the corresponding raw image, with multi-angle illumination provided by sequential activation. (d) Laser and motorized scanning mirror combination light source and the corresponding raw image, where illumination is provided by multi-angle scanning using the motorized scanning mirror.}
	\label{Fig.1}
\end{figure}

One conventional strategy to alleviate ill-posedness is to approximate the forward physical model\textsuperscript{\cite{kamm2013inversion,brown1967validity}}. Drawing on the weakly scattering approximation (1st Born or Rytov)\textsuperscript{\cite{wolf1969three,gbur2002diffraction}}, Wolf et al. introduced a single-scattering model (Fig.\ref{Fig.3}(a)), simplifying the inverse problem\textsuperscript{\cite{beydoun1988first}}. Through filtered back propagation and phase retrieval\textsuperscript{\cite{zuo2020wide}}, researchers reconstructed the 3D scattering potential from 2D intensity stacks measured at different illumination angles, while Kramers–Kronig relations further separated absorption and phase contributions under certain conditions\textsuperscript{\cite{baek2021intensity}}. Our previous work employed a circular multi-angle illumination scheme to linearize the inverse problem, thereby enabling direct analytical RI reconstruction\textsuperscript{\cite{li2019high}}. Nonetheless, such approximations fail for thicker or highly scattering specimens\textsuperscript{\cite{chen2020multi}}, and single-scattering approaches inherently suffer from the missing cone problem\textsuperscript{\cite{lim2015comparative}}.

To accommodate multiple scattering, researchers introduced the multi-slice beam propagation (MSBP) model\textsuperscript{\cite{chowdhury2019high}}, adapted from multi-layer seismic exploration methods\textsuperscript{\cite{alaei2012seismic}} (Fig.\ref{Fig.3}(a)). Combining MSBP with alternating projection (AP) (referred to as MSBP-AP) significantly improves reconstruction accuracy by performing spatial-domain updates, thus mitigating the missing cone issue\textsuperscript{\cite{kamilov2015learning}}. However, strong dependence on oblique/off-axis illumination often yields artifacts (Fig.\ref{Fig.2}(c), Fig.\ref{Fig.3}(b)), degrading axial resolution and quantitative accuracy. Additionally, alternating projection can succumb to optimization collapse, failing to converge robustly.

\begin{figure}
	\centering
	\includegraphics[scale=0.32]{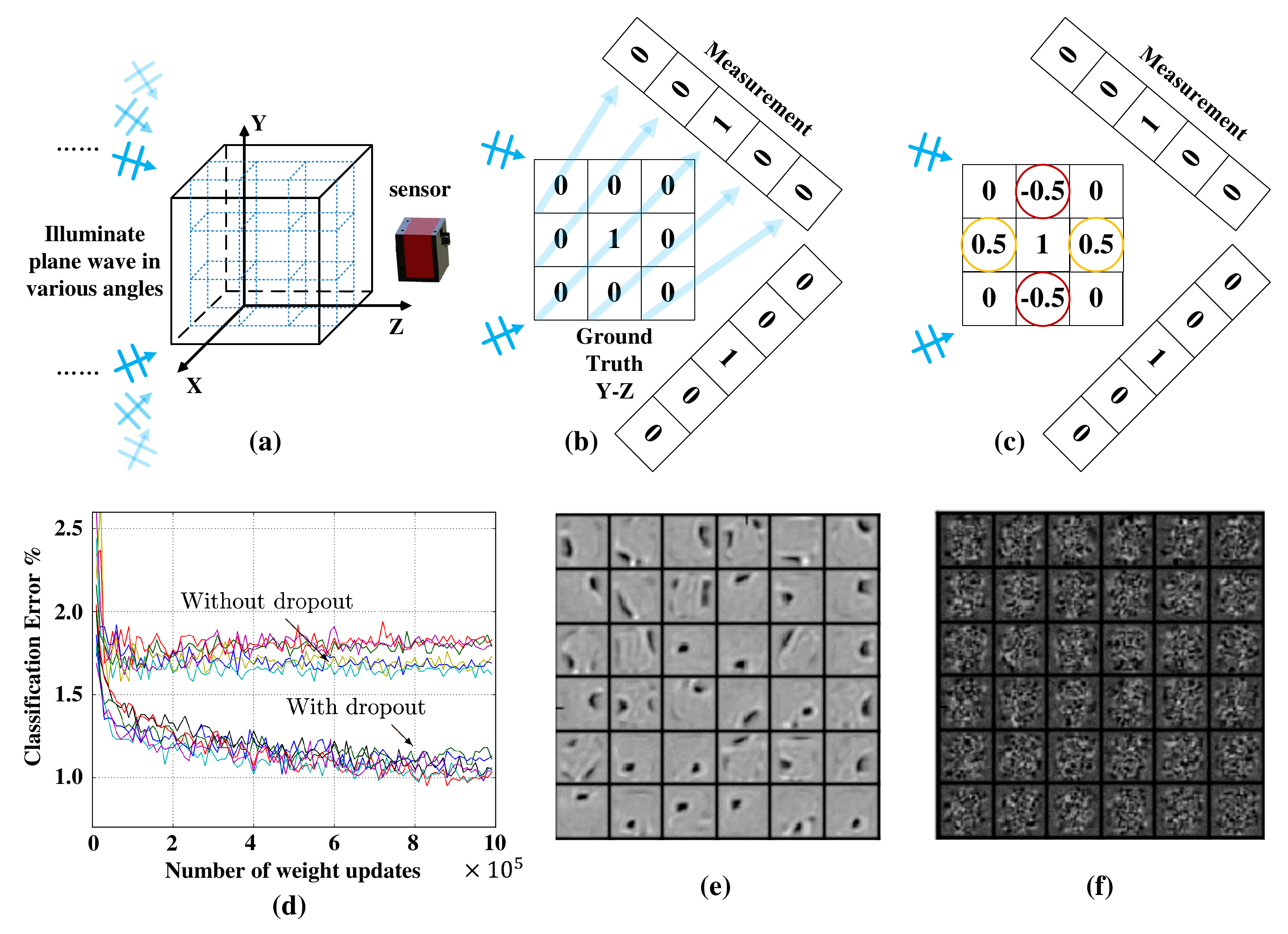}
	\caption{(a) Simplified imaging model of the non-interference microscopy system. (b)(c) The imaging process is further simplified to an accumulation along a single direction, with observations constrained to the y-z plane of the 3D sample. Due to the pronounced effect of oblique illumination, the same intensity images are reconstructed as local optimal solutions containing artifacts. The red and yellow circles highlight the reconstructed negative artifacts and axial artifacts, respectively. (d) Test error for various architectures, comparing results with and without dropout. (e) Visualization results of a subset of convolutional kernels from the hidden layer, comparing the effect of using and not using dropout.}
	\label{Fig.2}
\end{figure}

In recent years, the accelerating field of AI for science has enabled deep learning methods to effectively tackle nonlinear ill-posed problems. In label-free 3D biological imaging, researchers have explored data-driven strategies to learn nonlinear mappings from measurements to reconstructions\textsuperscript{\cite{matlock2023multiple}} or employed Physics-Informed Neural Networks (PINNs) and Deep Image Prior (DIP)\textsuperscript{\cite{liu2022recovery}}—leveraging convolutional neural network (CNN) biases—to address the ill-posedness\textsuperscript{\cite{barbastathis2019use}}. However, these approaches typically focus on weakly scattering samples and often grapple with result confidence issues or exhaustive parameter searches.

In this work, we pursue a Differentiable Imaging strategy to rigorously embed the physics of light propagation within a neural optimization framework. Instead of using a pre-trained model to map raw measurements to reconstructions, we treat the sample itself as a neural network, where each voxel’s complex value represents the complex amplitude transmittance at that spatial position (Fig.\ref{Fig.3}(b))\textsuperscript{\cite{yang2023refractive,yang2022fourier}}. This ensures that each forward pass is an accurate, end-to-end differentiable simulation of the optical system, preserving physical fidelity.

Nevertheless, realizing high-quality reconstructions demands attention to certain key aspects:
\begin{itemize}
	\item The conventional MSBP-based method (MSBP-AP)—a state-of-the-art solution for intensity-only tomography—often fails with complex samples, risking collapse or convergence to poor solutions. In contrast, advanced neural optimization strategies can better explore the parameter space\textsuperscript{\cite{dosovitskiy2020image,han2022survey}}.
	
	\item MSBP-AP reconstructions frequently suffer from background artifacts, where supposed background regions exhibit RI values exceeding the medium’s actual RI. This effect resembles neuron over-activation in neural networks lacking dropout\textsuperscript{\cite{hinton2012improving}}.
	
	\item In the absence of dropout, excessive reliance on certain neurons leads to co-adaptation and non-ideal feature patterns (Fig.\ref{Fig.2}(d)), whereas dropout encourages sparse and interpretable features, reflected in Fig.\ref{Fig.2}(e)\textsuperscript{\cite{srivastava2014dropout}}.
\end{itemize}

\begin{figure*}
	\centering
	\includegraphics[scale=0.17]{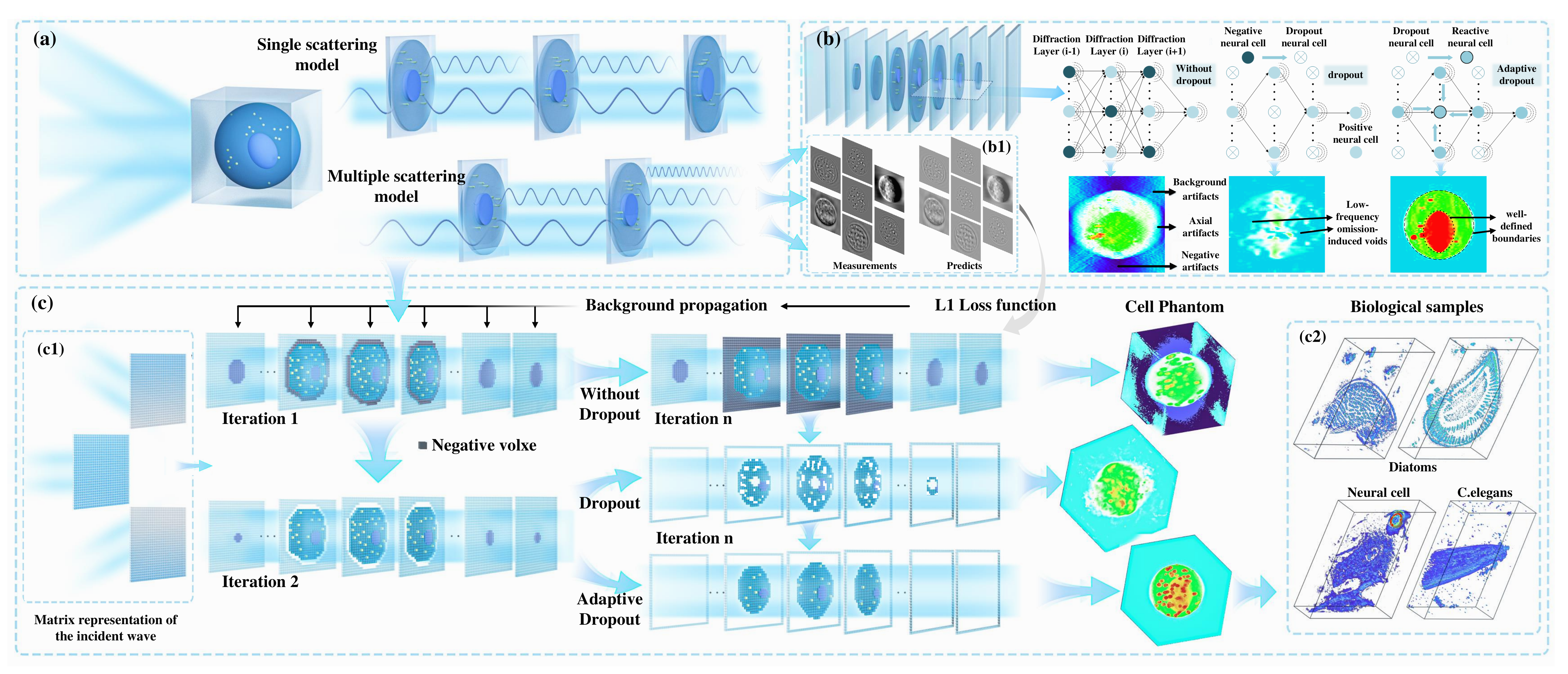}
	\caption{(a) Two typical light propagation models through a 3D sample: the single scattering model and the multiple scattering model. (b) Representation of the 3D sample as an ANN, demonstrating the implementation of adaptive dropout within the network. (b1) Predicted intensity images from ADNN based on the multiple scattering model (MSBP in this work) compared to the actual measured intensity images during optimization. (c) Training processes of the ADNN, where the L1 difference between predicted and measured values guides the optimization. This panel shows the reconstruction of a cell phantom under three scenarios: without dropout, with dropout, and with adaptive dropout. (c1) Matrix representation of the incident wave used as input to ADNN. (c2) Visualization of the 4D point cloud representing the reconstructed biological sample post-application of adaptive dropout.}
	\label{Fig.3}
\end{figure*}

While conventional artificial neural networks (ANNs) emphasize global input–output mappings, our differentiable programming paradigm demands careful voxel-wise accuracy and rigorous physical constraints, rendering standard random dropout suboptimal for multi-layer physical neural networks.

To address these issues, we propose a physics-based adaptive dropout method. Exploiting the prior that the sample’s RI typically exceeds that of the surrounding medium\textsuperscript{\cite{khan2021refractive}}, we remove neurons corresponding to negative RI voxels during optimization and selectively reactivate them in neighboring regions when warranted. We term this network the Adaptive Dropout Neural Network (ADNN) to emphasize its ability to dynamically regulate neuron activity based on the local spatial context of the sample. As illustrated in Fig.\ref{Fig.3}(c), we compare the optimization process and reconstructed results obtained with no dropout, standard dropout, and adaptive dropout, demonstrating clear advantages in convergence behavior and final image quality.

\section*{\textbf{Results}}
\subsection*{\textbf{\small Quantitative experimental validation}}
To validate the 3D quantitative RI reconstruction capability of ADNN, we conducted both the simulated and real experiments on phase-only polystyrene microspheres with the same parameters (8$\upmu$m diameter, RI=1.60). We constructed two specific experiments, in which the microsphere were immersed in RI matching media with indices of $n_{m}=1.56$ and $n_{m}=1.58$, respectively, to evaluate the reconstruction under different scattering conditions. To achieve this, an LED matrix was employed as the illumination source, providing varied illumination angles, and two representative raw images are presented in Fig.\ref{Fig.1}(i)(ii).

To fairly compare the tomographic performance of our proposed method with that of MSBP-AP, we incorporated non-negativity regularization into MSBP-AP —— a concise and effective approach to incorporate physical priors about the refractive index distribution (i.e., the sample’s RI is generally higher than that of the background medium)

From the reconstructed results shown in Fig.\ref{Fig.4}(a*)-(h*), it is evident that the MSBP-AP method suffers from severe axial artifacts, leading to elongation of axial features and reduced axial resolution, as well as prominent background artifacts. In contrast, the ADNN reconstruction results are largely free from artifacts, maintaining significantly higher axial resolution compared to MSBP-AP. Overall, the signal-to-noise ratio (SNR) in the 3D reconstructions using ADNN, both in simulations and real experiments, exhibits a marked improvement over MSBP-AP.

In addition, as illustrated in Fig.\ref{Fig.4}(*1)-(*3), the refractive index (RI) distribution curves indicate that ADNN's measurements are more consistent, approaching RI values of 0.02 or 0.04. Additionally, the one-dimensional (1D) RI profile along the orange line aligns more closely with the true distribution (which resembles a single square wave), underscoring the superior fidelity of the ADNN reconstruction.

Furthermore, Fig.\ref{Fig.4}(i)(j) presents quantitative evaluations of the simulation results, demonstrating significant improvements in SSIM and Peak Signal-to-Noise Ratio (PSNR) for ADNN over MSBP. The MAE and L1 loss curves in Fig.\ref{Fig.4} indicate that the ADNN reconstructed results not only converge faster but also progressively approach the true values, whereas the MSBP fails to converge accurately.

\begin{figure*}
	\centering
	\includegraphics[scale=0.23]{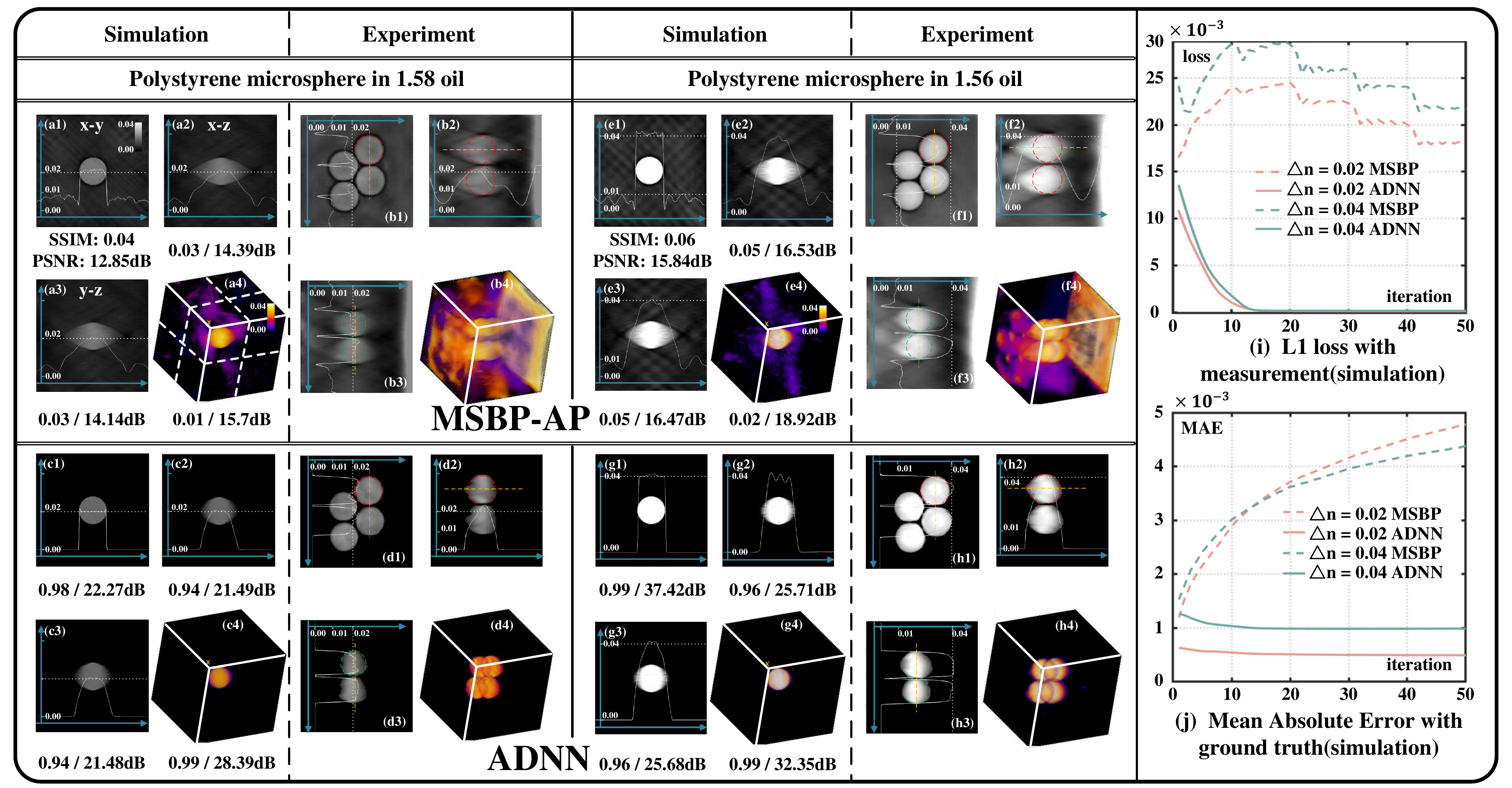}
	\caption{The tomographic reconstructed results of phase-only sample (polystyrene microspheres). The illumination was provided by an LED matrix illuminator. A RI difference of 0.02 represents weakly scattering samples, while a RI difference of 0.04 characterizes multiple scattering samples. (*1) Slice of polystyrene microspheres in the x-y plane at z = 0$\upmu$m. (*2) Slice of microspheres in the x-z plane at y = 16.25$\upmu$m. (*3) Slice of microspheres in the y-z plane at x = 16.25$\upmu$m. The planes represented by (*1), (*2), and (*3) correspond to the white dashed planes shown in (a4). In the experiment, the coordinate axes and white curves in (*1), (*2), and (*3) represent the 1D RI distribution along the orange dashed path. For the simulation, they represent the distribution along the central path. (*4) 3D maximum intensity projection of microspheres with auto-modulation display\textsuperscript{\cite{long2012visualization}}. (i) Variation curve of the L1 loss between predicted and measured values during optimization. (j) Variation curve of the MAE between reconstructed results and true values throughout optimization.}
	\label{Fig.4}
\end{figure*}

\subsection*{\textbf{\small Experimental validation on weakly-scattering biological sample}} 
In many transmission microscopy observations, a large portion of samples are weakly scattering and transparent. The imaging performance on these types of samples is crucial for evaluating the effectiveness of imaging methods. We chose three typical biological samples —— rat hippocampal neuronal cells, representing an animal cell, and two types of algal cells, diatom microalgae (S68786,Fisher Scientific) and Surirella spiralis (S.spiralis) diatom —— to demonstrate the effectiveness of the proposed method for weakly scattering samples. 

Among these samples, the rat hippocampal neuronal cells adhere to the culture dish and have relatively small axial dimensions. The additionally selected two types of diatoms are also very thin and exhibit a 3D distribution. The imaging performance on these three samples can significantly represent the imaging effects on weakly scattering samples.

In the experiment using rat hippocampal neuronal cells, we employed an LED matrix as the illumination source. Two example measurements are provided in Fig.\ref{Fig.1}(iii)(iv). For the reconstruction method, we used reconstruction method based on the 1st Born approximation (hereafter referred to as the 1st Born) and MSBP-AP as baseline methods for comparison, as both methods have been extensively validated under similar imaging settings. 

In addition, If the 1st Born method were combined with non-negativity regularization, it would lead to a significant degradation in the final reconstruction quality.  Therefore, non-negativity regularization was not applied to the 1st Born approach. Alternatively, displaying only the non-negative portion of the reconstruction would not adhere to the underlying physical model on which the reconstruction is based (The non-negative display results are still provided in the Supplementary Information for reference.).

\begin{figure*}
	\centering
	\includegraphics[scale=0.24]{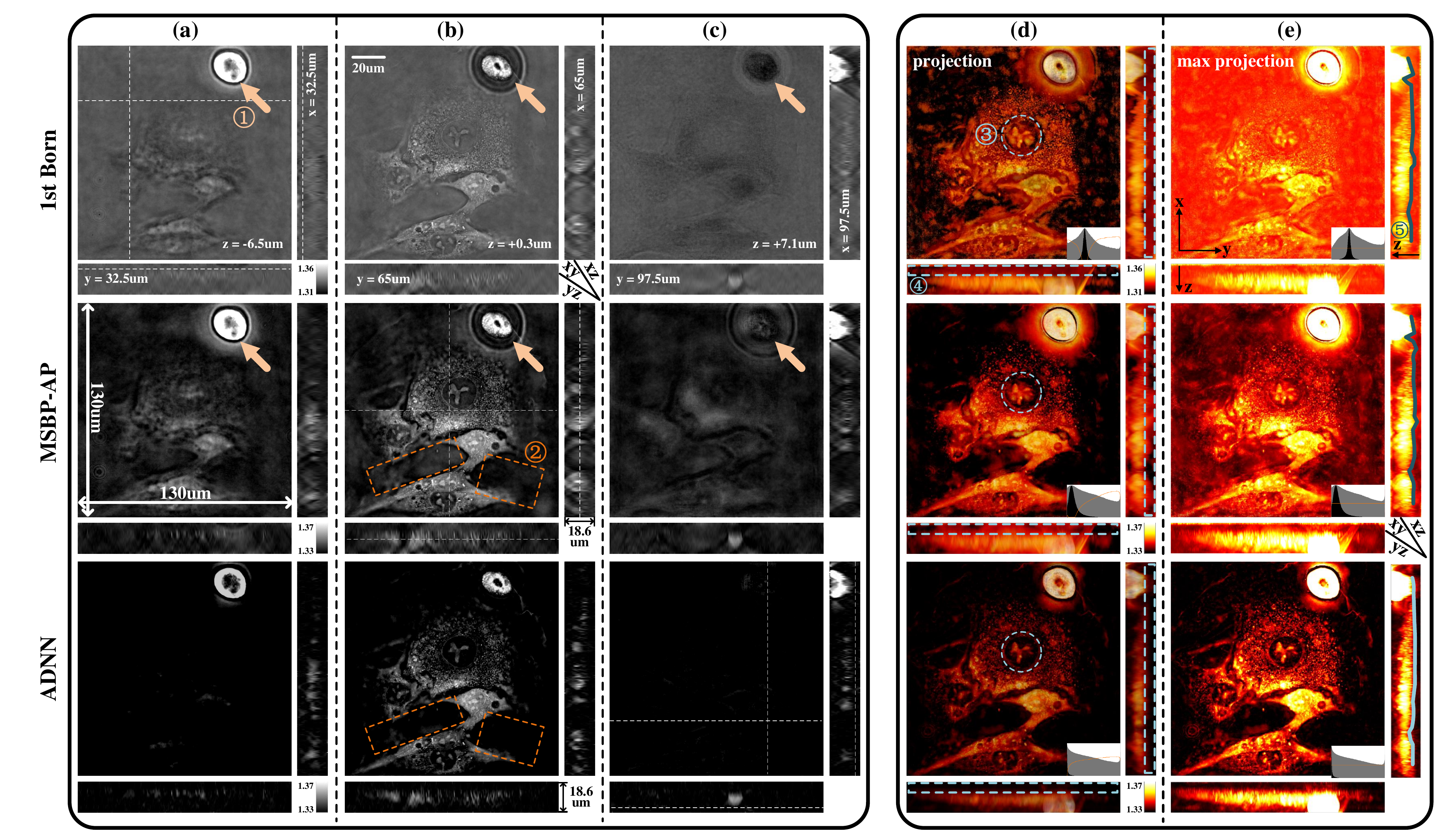}
	\caption{Tomographic reconstruction results of rat hippocampal neuronal cells (weakly scattering sample, illumination was provided by an LED matrix illuminator using partially coherent light). 
		(a) Reconstructed results at z = -6.5$\upmu$m in the x-y plane, at x = 32.5$\upmu$m in the y-z plane, and at y = 32.5$\upmu$m in the x-z plane, for different methods. 
		(b) Reconstructed results at z = +0.3$\upmu$m in the x-y plane, at x = 65$\upmu$m in the y-z plane, and at y = 65$\upmu$m in the x-z plane. 
		(c) Reconstructed results at z = +7.1$\upmu$m in the x-y plane, at x = 97.5$\upmu$m in the y-z plane, and at y = 97.5$\upmu$m in the x-z plane. 
		(d) 3D projection results generated in ImageJ with auto-modulation display. 
		(e) 3D max projection results without modulation display, with ID-TF curve from volume viewer in the bottom right corner.}
	\label{Fig.5}
\end{figure*}

The reconstructed results, as illustrated in Fig.\ref{Fig.5}, show the imaging performance. In the results obtained using the 1st Born and the MSBP-AP, artifacts were observed within the nuclei and along the edges of adherent cells, as well as around cells (as indicated by part \textcircled{1}) that had detached from the culture dish wall post-mortem. Furthermore, the background in the MSBP-AP was non-uniform, exhibiting background artifacts (as shown in part \textcircled{2}\textcircled{4}) and artifacts within the nuclei(part \textcircled{3}). In constrast, the results reconstructed using the ADNN, as observed in the tomographic results, exhibited minimal artifacts.

The axial dimensions were more closely aligned with the lateral dimensions, reflecting a more accurate depiction of the actual distribution. This aligns with the enhancements in axial resolution demonstrated by the ADNN in both simulation and the quantitative experiments using microspheres discussed in the main text. In conclusion, these improvements confirm that the ADNN effectively enhances optical-sectioning capabilities while suppressing artifacts, thereby providing a clearer and more accurate representation of the actual structure.

In the 3D projection results of the ADNN method, the neuron information is prominently highlighted with minimal artifacts. The projection results in the xz and yz directions clearly show that the cells adhere to a specific xy plane, closely resembling the growth morphology of cells on the culture dish (part \textcircled{5}).

To eliminate the potential impact of One-Dimensional Transfer Function(1D-TF) display modulation on the final results, we further present a comprehensive analysis of the projection display effects under different 1D-TF display modulations for these methods in the Supplementary Information.

\begin{figure*}[htb]
	\centering
	\includegraphics[scale=0.13]{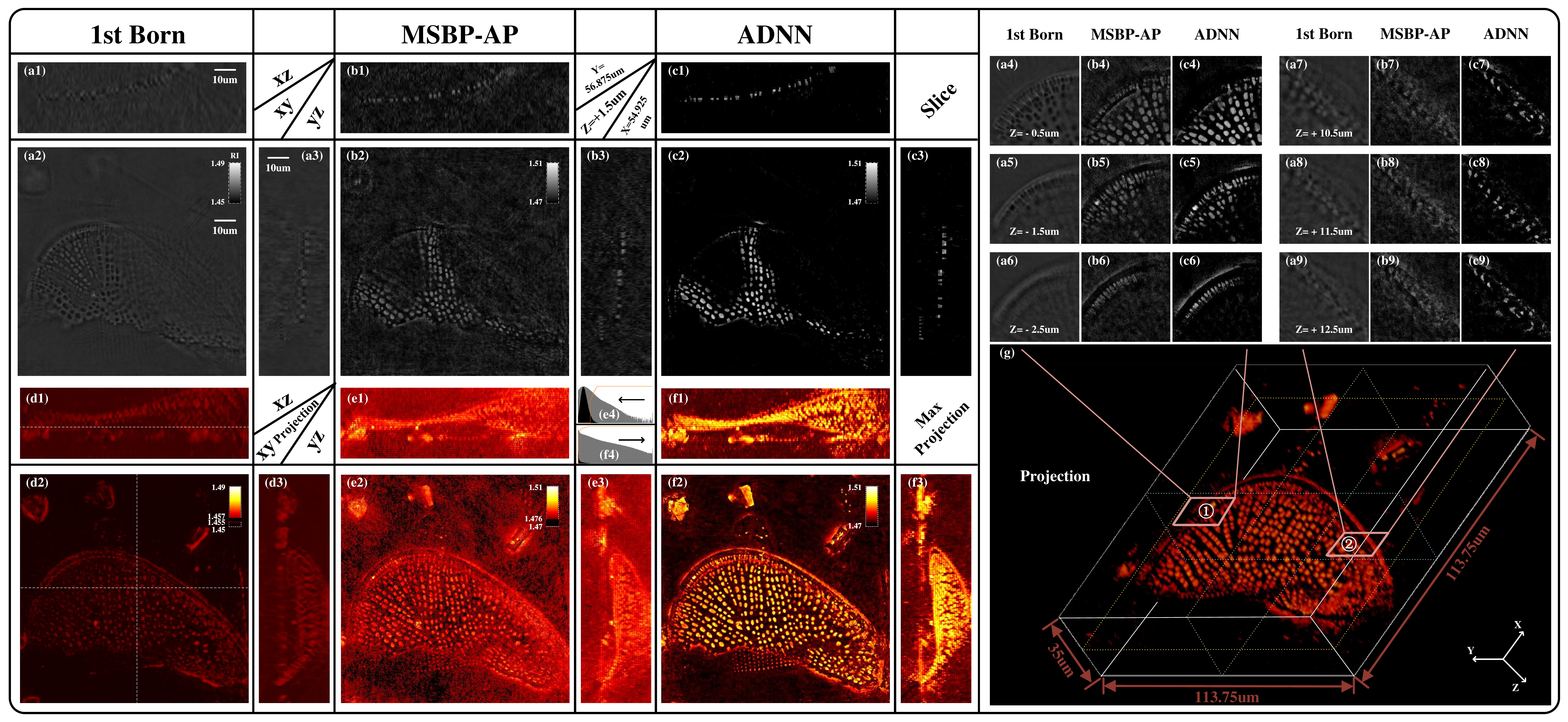}
	\caption{The tomographic reconstructed results of diatom microalgae (weakly scattering samples, illumination was provided by an LED ring illuminator using partially coherent light).
		(a1), (b1), (c1) The reconstructed results at y = 56.875 $\upmu$m in the x-z plane using the 1st Born, MSBP-AP, and ADNN, respectively.
		(a2), (b2), (c2) The reconstructed results at z = +1.5 $\upmu$m in the x-y plane using different reconstruction methods.
		(a3), (b3), (c3) The Reconstructed results at x = 54.925$\upmu$m in the y-z plane.
		(*4) - (*6) The reconstructed results at region \textcircled{1} for z = -0.5 $\upmu$m, z = -1.5 $\upmu$m, and z = -2.5 $\upmu$m in the x-y plane using different methods.
		(*7) - (*9) The reconstructed results at region \textcircled{2} for z = +10.5$\upmu$m, z = +11.5$\upmu$m, and z = +12.5$\upmu$m in the x-y plane using different methods.
		(d1)-(d3) 3D projection results generated in ImageJ with auto-modulation display, using 1st born.
		(e1)-(e3) 3D max projection results generated in ImageJ with 1D-TF(e4) display modulation, using MSBP-AP.
		(f1)-(f3) 3D max projection results generated in ImageJ with 1D-TF(f4) display modulation, using ADNN.
		(g) 3D projection results generated in ImageJ with auto-modulation display, using ADNN.}
	\label{Fig.6}
\end{figure*}

We subsequently employed annular illumination —— the LED ring illuminator to explore the efficiency of data processing with ADNN and to assess the quality of 3D reconstruction under these conditions. In the experiment using diatom algae as the sample, we collected 24 images within 7 seconds. Two example measurements are provided in Fig.\ref{Fig.1}(v)(vi). Despite the sample is transparent (indicating it has low-absorbing characteristic that renders it nearly invisible under bright-field illumination), its optical feature can modulate the incident plane wave reveals the phase features of the sample, which become apparent due to the asymmetric illumination. Fig.\ref{Fig.6} presents the comparison results of the 3D tomography slice and max projection in various directional planes.

Based on our prior knowledge of the sample, the diatom microalgae which fixed in glycerin gelatin is a unicellular algae with relatively clear borders and regular arrangement of punctae. The sample, although dispersed within a 3D space in the medium, is very thin, a fact corroborated by the reconstructed results. Consequently, the internal scattering within the sample is notably weak.

However, the 1st Born fails to eliminate light from regions above and below the plane of interest, as depicted in Fig.\ref{Fig.6}(a2), (a4)-(a6). This limitation leads to the appearance of a RI in the reconstruction that is lower than that of glycerin gelatin, which is attributable to the effect of the defocused light field. As illustrated in Fig.\ref{Fig.6}(d1),(d3), the projection results of the 1st Born show that the sample appears elongated along the z-axis direction, a phenomenon also caused by the limitation of the 1st Born.

The limitations of the 1st Born are significantly improved by the MSBP-AP, as demonstrated in the results shown in the corresponding column of Fig.\ref{Fig.6}.
Specifically, the reconstructed results in (b1)-(b3) demonstrate that no negative RI values (indicative of erroneous reconstructions) are present. However, MSBP-AP inadvertently alters the RI values in the background, resulting in a relatively low SNR in the final images. 

The results obtained using ADNN are presented in the corresponding column of Fig.\ref{Fig.6}. Experimental findings demonstrate that ADNN effectively mitigates the influence of light fields originating from above and below the plane of interest, significantly reducing artifacts and confining the accurate refractive index information to its correct spatial location. This artifact reduction leads to the elimination of interference from artifact-related RIs on light field modulation, resulting in enhanced quantitative 3D RI reconstruction. As shown in Fig.\ref{Fig.6}(c1)-(c9), the reconstructed RI using ADNN closely aligns with the actual RI of the diatom, which is 1.51, indicating superior reconstruction accuracy compared to other methods. Additionally, as shown in Fig. \ref{Fig.6} (*4)-(*9), ADNN demonstrates clearer reconstructed results in certain regions, further showcasing its improved performance over other approaches.

The high SNR in the imaging and clearest 3D projection results further confirms the effectiveness and superiority of ADNN. More detailed tomographic scan results and angle-by-angle projection results are presented in Supplementary Videos.

\begin{figure*}[htb]
	\centering
	\includegraphics[scale=0.13]{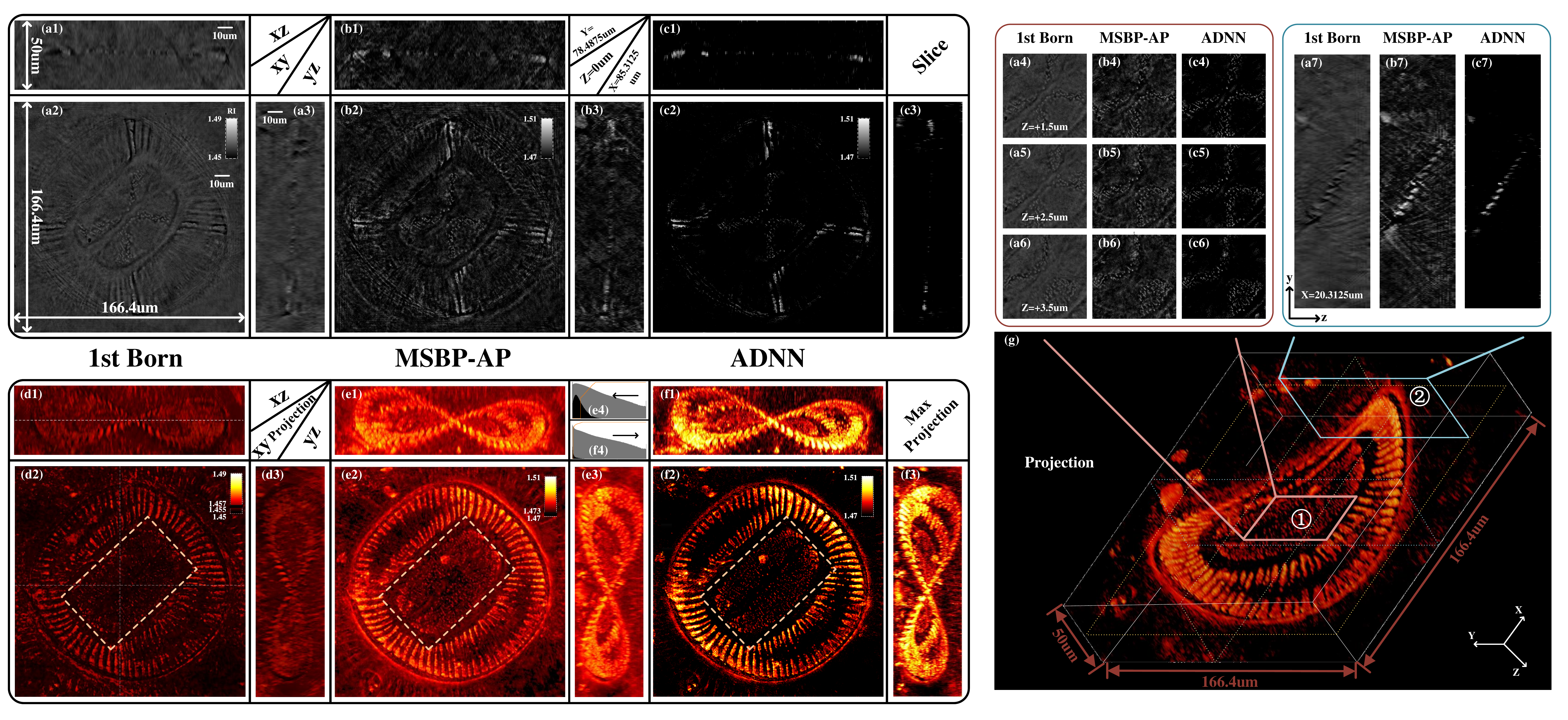}
	\caption{The tomographic reconstructed results of S.spiralis diatom (weakly scattering samples, illumination was provided by an LED ring illuminator using partially coherent light). 
		(a1) - (c1) The reconstructed results at y = 78.4875 $\upmu$m in the x-z plane using the 1st Born, MSBP-AP, and ADNN, respectively.
		(a2) - (c2) The reconstructed results at z = 0 $\upmu$m in the x-y plane using different reconstruction algorithms.
		(a3) - (c3) The Reconstructed results at x = 85.1325$\upmu$m in the y-z plane.
		(*4) - (*6) The reconstructed results at region \textcircled{1} for z = +1.5 $\upmu$m, z = +2.5 $\upmu$m, and z = +3.5 $\upmu$m in the x-y plane using different reconstruction algorithms.
		(a7) - (c7) The reconstructed results at region \textcircled{2} for x = 20.3125 $\upmu$m in the y-z plane.
		(d1)-(d3) 3D projection results generated in ImageJ with auto-modulation display, using 1st born.
		(e1)-(e3) 3D max projection results generated in ImageJ with 1D-TF(e4) display modulation, using MSBP-AP.
		(f1)-(f3) 3D max projection results generated in ImageJ with 1D-TF(f4) display modulation, using ADNN.
		(g) 3D projection results generated in ImageJ with auto-modulation display, using ADNN.}
	\label{Fig.7}
\end{figure*}

The Surirella spiralis (S.spiralis) diatom possesses optical properties similar to those of the previously mentioned diatom microalgae sample; however, it is morphologically distinct, characterized by its spiral form, as indicated by its name. We also used the LED ring for illumination in the experiment with S.spiralis. Two example intensity measurements are presented in Fig.\ref{Fig.1}(vii)(viii). The imaging results, shown in Fig.\ref{Fig.7}, clearly reveal the distinct helical structure from the projection results on the x-z and y-z planes. 

The reconstructed results from various methods further substantiate the advantages of the ADNN. From the comparative results shown in Fig.\ref{Fig.7}(*1)-(*7), it is evident that ADNN effectively eliminates the majority of background artifacts while preserving the accurate refractive index information of the biological sample. Regarding the detailed structural information, as observed in (d2), (e2), and (f2), the high-frequency components at the center of the S. spiralis are prominently preserved. This observation is further corroborated in the comparisons presented in (*4)-(*6).

Notably, in displaying the 3D projection results of the MSBP-AP, the 1D-TF of the 3D projection of the MSBP-AP reconstructed results was adjusted meticulously. The 1D-TF controls the projection brightness across different RIs on the observation plane and selects the maximum RI for projection along a path perpendicular to the observation plane, effectively hiding noise information based on an adaptive threshold. This adjustment ensures that artifacts are minimized and sample information is preserved. However, such stringent adjustments are unnecessary for the ADNN reconstructions. In the experiments of this chapter, the optimally post-processed MSBP-AP projection was compared with the minimally post-processed ADNN projection. Despite the careful adjustments to the 1D-TF of the MSBP-AP, the imaging quality of the reconstructions based on the alternating projection method remains inferior to that of ADNN.

Moreover, this compromise between artifact removal and information retention is detrimental to quantitative microscopic analysis. Since it is challenging to accurately differentiate between artifacts and genuine informational content, any artificial manipulation of display results might conceal critical data. Further artifact removal in alternating projection risks losing critical high-frequency information at the center of the diatom algae (details in Supplementary Information). These observations further highlight the benefits of ADNN reconstructions, including more accurate RI, higher SNR, and improved user-friendliness in 3D displays.

\subsection*{\textbf{\small Experimental validation on multiple-scattering biological sample-Caenorhabditis elegans(C.elegans)}}

\begin{figure*}[htb]
	\centering
	\includegraphics[scale=0.175]{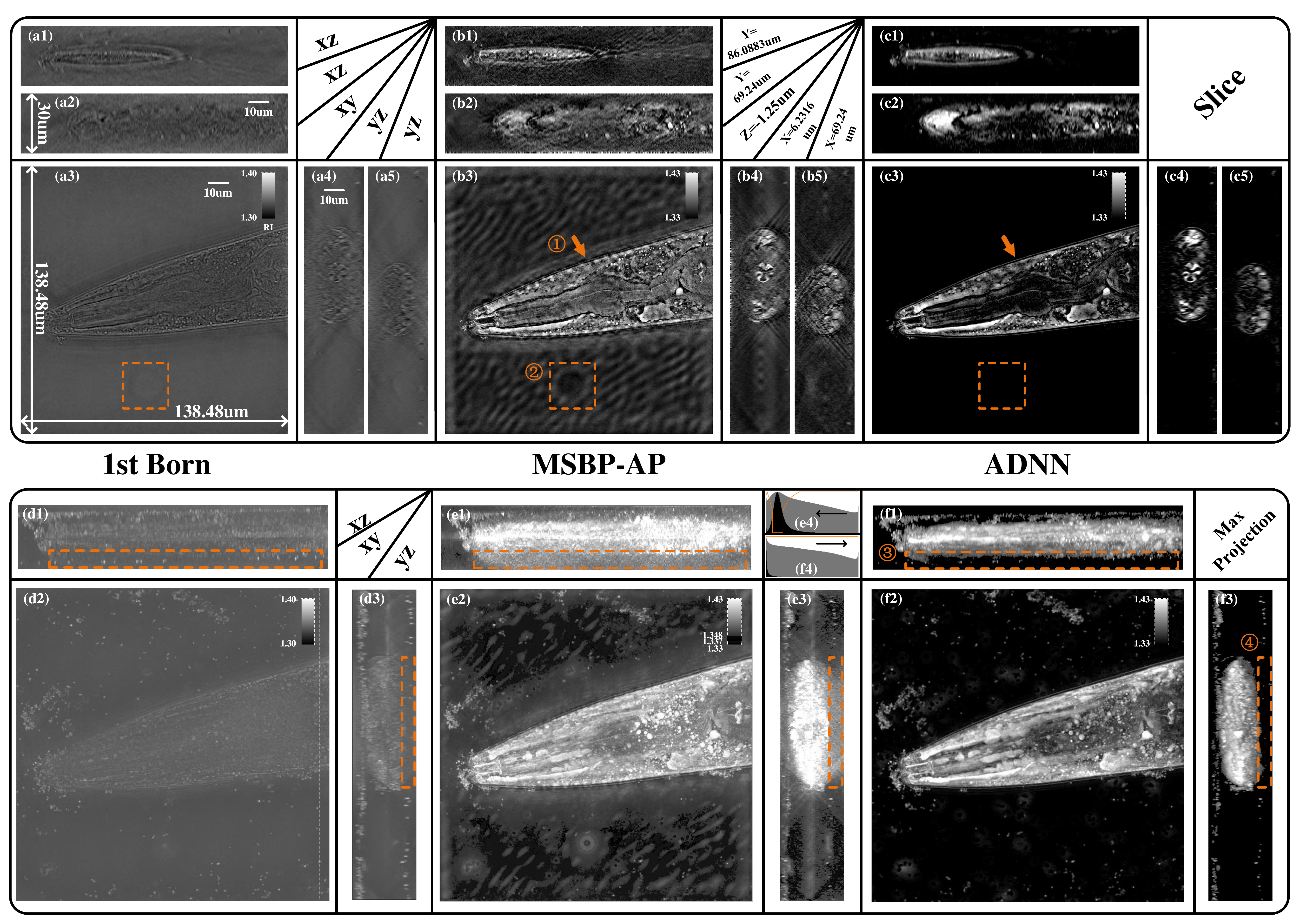}
	\caption{The tomographic reconstructed results of C.elegans (multiple scattering sample, Illumination was provided by a coherent illumination system —— motorized mirror mount system). 
		(a1) - (c1) The reconstructed results at y = 86.0883 $\upmu$m in the x-z plane, obtained using the 1st Born, MSBP-AP, and ADNN, respectively
		(a2) - (c2) The reconstructed results at y = 69.24 $\upmu$m in the x-z plane using different algorithms.
		(a3) - (c3) The reconstructed results at x = -1.25 $\upmu$m in the x-y plane.
		(a4) - (c4) The reconstructed results at x = 6.2316 $\upmu$m in the y-z plane.
		(a5) - (c5) The reconstructed results at x = 69.24 $\upmu$m in the y-z plane.
		(d1) - (d3) 3D projection results generated in ImageJ with auto-modulation display, using 1st born.
		(e1) - (e3) 3D max projection results generated in ImageJ with 1D-TF(e4) display modulation, using MSBP-AP.
		(f1) - (f3) 3D max projection results generated in ImageJ with 1D-TF(f4) display modulation, using ADNN.
	}
	\label{Fig.8}
\end{figure*}

Apart from weakly scattering samples, a significant portion of biological samples, although transparent, exhibit multiple scattering effects due to their increased thickness and complex internal structures, resulting in a semi-transparent appearance. To validate the imaging performance of ADNN for samples with multiple scattering effects, we selected C.elegans, a model organism in life sciences, as a representative sample for our experiments.

The raw data of C.elegan were obtained from Laura Waller's Computational Imaging Lab at UC Berkeley\textsuperscript{\cite{chowdhury2019high}}.
Similar to the previous experiment, the raw data were collected using multi-angle plane wave illumination of the sample. However, the system utilized for acquiring this raw data differs in that it employs motorized mirror mount(MMM) to achieve the multi-angle illumination.

Fig.\ref{Fig.1}(x)(xii) presents two examples of total field intensity measurements. The raw data for the mouth and pharynx, illuminated by positive incidence, are shown in Fig.\ref{Fig.1}(ix)(xi). Unlike measurements of weakly scattering samples such as diatoms, where information on the focal plane can still be discerned, measurements from C.elegans typically yield little discernible information. This lack of clarity is attributed to internal multiple-scattering within the C.elegans. Shwetadwip Chowdhury et al. have previously analyzed such conditions. Many biological samples, similar to the C.elegans, exhibit multiple-scattering; they are generally thicker and possess complex internal structures, making it nearly impossible to discern any information at a single focal plane. However, it is precisely these types of samples that are in greater need of tomographic imaging, which allows for a clearer visualization of their internal structures.

The first column of Fig.\ref{Fig.8} shows 1st Born the reconstructed results. It is challenging to discern the internal structure of C.elegans from these results, due to the severe degradation of the method when the weak scattering approximation (which neglects multiple scattering within the sample) no longer holds for C.elegans.

The second column of Fig.\ref{Fig.8} presents the reconstructed results of MSBP-AP. Although the reconstruction quality is significantly improved compared to the 1st Born severe artifacts are evident. These artifacts are concentrated around the background and edges of C.elegans in the x-y plane, resulting in a cluttered background and blurred membrane structures at C. elegans' boundary (part \textcircled{1}). Furthermore, due to the strong scattering effects, cross-talk occurs between planes at different axial positions(part \textcircled{2}), causing additional degradation of the reconstruction quality.

The artifacts in the x-z and y-z planes of the MSBP-AP reconstructed results are more pronounced, showing a similar elongation effect as observed in microspheres during previous quantitative experiments. This is due to the larger lateral dimensions of C.elegans, which lead to more severe axial artifacts compared to weakly scattering samples (part \textcircled{3}, \textcircled{4}).

The reconstructed results of ADNN in the corresponding column of Fig.\ref{Fig.8} demonstrate the highest imaging quality, clearly outperforming other methods in terms of accuracy and artifact reduction. The background is exceptionally clean, and the boundaries of C.elegans are very distinct, with no axial artifacts present. Moreover, it does not increase the data acquisition or reconstruction time compared to the 1st Born and MSBP-AP, but exhibiting outstanding performance.

\section*{\textbf{Disscussion}}
To provide a comprehensive discussion of our proposed method, we first focus on improvements made to the forward model. Our multi-layer network's forward propagation method fully incorporates the physical processes of multiple scattering within the sample, providing more accurate and complete results compared to single scattering models. Additionally, instead of using a conventional multi-slice model with an analytical beam propagation process to connect adjacent layers, we leverage differentiable programming, resulting in a more concise and efficient forward propagation across layers.

Furthermore, the physics-guided adaptive dropout optimization algorithm presented in this work integrates the strengths of analytical solutions based on physical information and optimization methods based on deep learning. In contrast to random dropout in classical ANNs, which primarily enhances network robustness without considering the physical meaning of individual parameters, our ADNN framework combines the benefits of stochastic deactivation with physics-based guidance. This alignment with physical laws ensures the network parameters are optimized in a manner that accurately reflects the underlying physical phenomena, thereby enhancing both scientific rigor and predictive accuracy.

Due to these innovations, the optical diffraction tomography reconstruction method we propose achieves high-fidelity 3D RI reconstructions for thick samples with strong internal scattering. It effectively mitigates issues such as axial artifacts, background artifacts, and numerical inaccuracies often encountered with traditional methods, including single scattering models(such as 1st born) and multiple scattering models(such as MSBP).

Another approach that integrates physical information with neural networks is to construct a loss function for neural network optimization based on physical processes. PINNs are example of this method. When applied to optical diffraction tomography, these methods also yield satisfactory reconstruction results, albeit with a reconstruction process that is considerably time-consuming.Typically, DIP-based reconstruction methods are very time-consuming; for instance, reconstructing a $1000\times1000\times40$ C.elegans(referencing DeCAF) requires approximately 20 hours.Compared to the aforementioned methods, the approach presented in this paper delivers similar or improved results with a reconstruction speed that is over 20 times enhanced.

Non-interference tomographic reconstruction of low-frequency samples poses a significant challenge, particularly in traditional optical diffraction tomography where the missing cone problem exemplifies the difficulty of reconstructing lateral low-frequency information axially. While our method significantly mitigates axial artifacts and the missing cone issue, it still performs less optimally with low-frequency samples exhibiting slowly structural distributions changes compared to those with rapidly varying features. In future studies, we aim to further optimize our approach to handle such low-frequency components more effectively. We believe that borrowing more network structure design and optimization strategies and techniques from the field of deep learning under the constraints of physical laws is an alternative and effective way of AI for science. 

\section*{\textbf{Conclusion}}
We present a novel tomographic 3D reconstruction method, ADNN, which achieves high-quality quantitative 3D reconstruction of RI distributions from intensity measurements. We extensively validated the superiority of ADNN through quantitative simulations and experiments on multiple biological samples using three different systems. The results demonstrate that ADNN eliminates artifacts in the reconstruction process without increasing optimization time, while preserving the detailed information in biological samples. Quantitative simulations showed that ADNN can reduce the MAE by 3 to 5 times and improve the reconstructed SSIM by approximately 4 to 30 times compared to the state-of-art method.

\section*{\textbf{Methods}}
\subsection*{\textbf{\small Experiment coded illumination setup}} 
\textbf{\small LED matrix.} The experimental setup, as shown in Fig.\ref{Fig.1}(b), includes a $15\times15$ LED matrix positioned 37.5 $mm$ from the sample plane for varied-angle illuminations. Each LED produces a plane wave (wavelength 623 nm, bandwidth 20 nm), and the LED matrix is controlled via an STM32 microcontroller. An sCMOS camera (PCO. Panda 4.2, 6.5 $\upmu$m pixel size) operates at 5 fps with 16-bit depth, acquiring 150 intensity images. With an illumination NA of 0.65, which is less than the NA of the $40\times/0.7$ objective, all images are brightfield.

\textbf{\small LED ring.} The experimental setup is illustrated in Fig.\ref{Fig.1}(c). We used the bright-field microscope with LED ring illumination unit which can provide tilted incident matching the objective NA. The radius of the ring LED unit is $30mm$. The LED ring is placed $35$ mm away from the sample, whose center is aligned with the optical axis of the microscope. Each LED approximately provides spatially coherent quasimonochromatic illumination with central wavelength $\lambda=515nm$ and $20nm$ bandwidth. The LED matrix is controlled by an microcontroller(Arduino Uno) and is synchronized with the camera to scan through the LEDs at at 3.3 fps. We captured 24 images using a 40$\times$ microscope objective(0.65 NA, CFI Plan Achro).

\textbf{\small Motorized Mirror Mount illumination system.} The experimental setup, as shown in Fig. 1(d), includes a green LED coupled to a 50 $\upmu$m multimode fiber. The collimated output is directed to a mirror on a motorized mount. A 4f-system, composed of an achromatic doublet and a condenser objective, conjugates the mirror plane to the sample for illumination scanning. Both condenser and imaging objectives are Nikon CFI Plan Apo Lambda $100\times$, NA 1.45. A final 4f-system de-magnifies the output onto a 20 MP sensor. For a more detailed description of the experimental setup, refer to the paper by Chowdhury et al.\textsuperscript{\cite{chowdhury2019high}}.

\subsection*{\textbf{\small Algorithm settings for different samples}}
\textbf{\small Polystyrene microspheres.} We configured each method to reconstruct $200$ slices, each consisting of $200\times200$ pixels, equally spaced between $-16.25\upmu$m and $+16.25\upmu$m. This configuration creates a volumetric representation of $32.5\times32.5\times32.5\upmu$$m^{3}$, with voxel size $0.1625\times0.1625\times0.16\upmu$$m^{3}$. 
For a volume consisting of $200\times200\times200$ voxels, the total time required to complete the tomographic reconstruction after 30 iterations was 6.5 minutes.

\textbf{\small Rat hippocampal neuronal cells} We configured each method to reconstruct $30$ slices, each consisting of $800\times800$ pixels, equally spaced between $-9.3 \upmu m$ and $+9.3 \upmu m$. This configuration creates a volumetric representation of $130\times130\times18.6 \upmu m^{3}$, with voxel size $0.1625\times0.1625\times0.62\upmu$$m^{3}$. For a volume consisting of $800\times800\times30$ voxels, the total time required to complete the tomographic reconstruction after 30 iterations was 18 minutes.

\textbf{\small Diatom microalgae (S68786, Fisher Scientific).} We configured each method to reconstruct $35$ slices, each consisting of $700\times700$ pixels, equally spaced between $-17.5 \upmu m$ and $+17.5 \upmu$m. This configuration creates a volumetric representation of $113.75\times113.75\times35 \upmu$$m^{3}$, with voxel size $0.1625\times0.1625\times1\upmu$$m^{3}$. For a volume consisting of $700\times700\times35$ voxels, the total time required to complete the tomographic reconstruction after 30 iterations was 10 minutes.

\textbf{\small S.spiralis diatom.} Each method were configured to reconstruct $50$ slices of $1024\times1024$ pixels equally spaced between $-25\upmu m$ and $+25\upmu m$, forming a volume of $166.4\times166.4\times50\upmu$$m^{3}$, with voxel size $0.1625\times0.1625\times1\upmu$$m^{3}$. For a volume consisting of $1024\times1024\times50$ voxels, the total time required to complete the tomographic reconstruction after 30 iterations was 30 minutes.

\textbf{\small C.elegans.} Each method were configured to reconstruct $120$ slices of $1200\times1200$ pixels equally spaced between $-15\upmu$m and $+15\upmu$$m$, forming a volume of $138.48\times138.48\times30\upmu$$m^{3}$, with voxel size $0.1625\times0.1625\times1\upmu$$m^{3}$. The total imaging time to complete 30 iterations of a $1200\times1200\times120$ voxels volume was 2.4 hours.

\subsection*{\textbf{\small ADNN framework}}
The structure of the Beam Propagation Model(BPM) significantly resembles the hierarchical structure of ANNs. Therefore, we first represent the segmentation of a series of layers of a 3D object using neural network layers. By employing a physical prior, namely the angular spectrum diffraction equation, to replace the convolution operation, we establish the connection for ADNN. In this manner, the layers of the optical neural network can be interpreted as information of the 3D sample, where the information at the m-th layer is represented as $L_{m}(r) = n_{3d}(r,m\Delta z)-n_{media}$ indicating the RI difference between the 3D sample and the media. Additionally, various optimizers designed for DNNs, such as Adam, can accelerate the rapid convergence to the global optimum. Mathematically, the diffraction propagation in ADNN can be recursively expressed as:

\begin{gather}
	P(r,\Delta z) =  exp(j2\pi\Delta z(\dfrac{n_{media}}{\lambda}^2-\Vert u\Vert^2)^{1/2})
	\label{Eq.(1)}
\end{gather}
\begin{gather}
	t_{m}(r,\Delta z) = exp(j2\pi\Delta z \dfrac{n_{3d}(r,m\Delta z)-n_{media}}{\lambda})
	\label{Eq.(2)}
\end{gather}
\begin{gather}
	U_{m+1}(r) = t_{m}(r,\Delta z) \cdot \mathscr{F}^{-1}\{{P(r,\Delta z) \cdot \mathscr{F}\{U_{m}(r)\}}\}
	\label{Eq.(3)}
\end{gather}
\[
\quad m=0,1 \cdots,M-1
\]
where $P(r,\Delta z)$ denotes the angular spectrum diffraction equation that propagates a light field by distance $\Delta z$, $r$ denotes the 2D spatial position vector, $u$ denotes the 2D spatial frequency space coordinates vector, $t_{m}(r,\Delta z)$ denotes the phase modulation by the $m$th layer, $U_{m}$ and $U_{m+1}$ are the input and output light field of $m$th layer in ADNN. And the entire sample is divided into $M$ layers.
The boundary condition to initialize the recursively Eq.(\ref{Eq.(3)}) is the incident plane wave illuminating the sample, $U_{0}^i(r)=exp(jk_{i}r)$ where $k_{illu}$ is the illumination wave vector at a particular angle. The physical process from the exit light field of the sample to the sensor plane can be expressed as:

\begin{gather}
	t(r,-\dfrac{M\Delta z}{2}) = C(u, k_0) \cdot P(r,-\dfrac{M\Delta z}{2})
	\label{Eq.(4)}
\end{gather}
\begin{gather}
	U_{predict}(r) = U_{M}(r) \cdot \mathscr{F}^{-1}\{t(r,-\dfrac{M\Delta z}{2})) \cdot \mathscr{F}\{U_{M}\}\}
	\label{Eq.(5)}
\end{gather}
\begin{gather}
	I_{predict}(r) = |U_{predict}(r)|^2
	\label{Eq.(6)}
\end{gather}

where $U_{M}(r)$ represents the exit electric field of the final layer of the 3D sample, $C(u, k_0)$ denotes the coherent transfer function of the system. The light field captured by the objective $U_{predict}(r)$ accounts for the accumulation of the diffraction and multiple-scattering processes that occurred during optical propagation through the ($-\dfrac{M\Delta z}{2}$,$\dfrac{M\Delta z}{2}$) around the focal plane. The term $I_{predict}(r)$ represents the intensity distribution captured by the camera. To provide direction for the optimization process, we define the loss function for the ADNN as follows:

\begin{gather}
	\mathcal{L}(n_{3d}^{(e,i)}) = \left| I_{predict}^{(e,i)}(r) - I_{measurement}^{(i)}(r) \right|
	\label{Eq.(7)}
\end{gather}
\[
\quad i = 1,2,\cdots,N
\]

\begin{gather}
	n_{3d}^{(e,i+1)} = n_{3d}^{(e,i)} - \alpha \nabla_{n_{3d}} \mathcal{L}(n_{3d}^{(e,i)})
	\label{Eq.(8)}
\end{gather}

where $n_{3d}^{(e,i+1)}$ represents the current 3D RI at epoch $e$ and for the 
$i$-th light, and \( \alpha \) represents the learning rate, and $\nabla_{n_{3d}} \mathcal{L}(n_{3d}^{(e, i)}) $ is the gradient of the loss function with respect to parameter \( n_{3d} \). In this formulation, each light's optimization is tracked by 
$i$, and after all $N$ lights are optimized, the epoch counter $e$ increments by one, signifying the completion of one epoch.

After establishing the physics-guided neural network, which incorporates the beam propagation model, we applied a prior assumption that the sample's RI exceeds that of the medium. Initially, dropout is applied to neurons linked with negative RI values during optimization. Later, dropped neurons are reactivated to recover low-frequency sample information. In the implementation of adaptive dropout, we employ two typical strategies to optimize network performance effectively.
\begin{itemize}
	\item 
	During the optimization of each illumination-intensity image pair, we apply dropout to neurons associated with negative refractive indices. After a certain period, we reactivate the dropped neurons to recover the low-frequency information of the sample. This process can be expressed as:
	\begin{gather}
		n_{3d}^{(e,i)}(r,m\Delta z) = \begin{cases} 
			0, \quad \text{if } n_{3d}^{(e,i)}(r,m\Delta z) < 0 \\
			n_{3d}^{(e,i)}(r,m\Delta z), \quad \text{otherwise}
		\end{cases}
		\label{Eq.(9)}
	\end{gather}
	\[
	\quad m=1,2 \cdots,M
	\]
	This method shows significant effectiveness in samples with abundant low-frequency components, such as cells or tissues, due to their smooth and gradual structural changes. However, it may perform slightly less effectively for samples like diatoms, which exhibit complex microstructures characterized by high-frequency features in the spatial frequency domain, compared to the next alternative approach.
\end{itemize}
\begin{itemize}
	\item For samples that exhibit complex microstructures characterized, ADNN typically converges quickly. In this case, after performing optimization without dropout for 3-10 epochs ($T$), where an epoch is defined as optimizing across all illumination-intensity image pairs, we then introduce dropout to neurons associated with negative refractive indices. Subsequently, at regular intervals during each iteration, these neurons are systematically dropped as illustrated by Eq.\ref{Eq.(9)}.
\end{itemize}

During ADNN training, neurons (voxels) with lower RI may be mistakenly dropped due to negative artifacts. To address this, we reactivate them using Total Variation (TV) regularization(details in Supplementary Information), guiding the reactivation to enhance low-frequency recovery and artifact suppression.
\begin{gather}
	n_{3d}^{(e,i)\text{TV}} = \mathcal{TV}\{n_{3d}^{(e,i)}\}
	\label{Eq.(11)}
\end{gather}
\begin{gather}
	n_{3d}^{(e,i)} = \begin{cases}
		n_{3d}^{(e,i)\text{TV}} & \text{if } n_{3d}^{(e,i)\text{TV}} > 0 \\
		0 & \text{otherwise}
	\end{cases}
	\label{Eq.(12)}
\end{gather}
The complete process for 3D intensity-based RI imaging with ADNN is summarized in Algorithm 1(Extend Data Fig.1).

\section*{\textbf{Code and data availability}}
The data and code used for reproducing the results in the manuscript is available at \href{https://github.com/yang980130/Enhancing-Optical-Diffraction-Tomography-with-Physics-Guided-Adaptive-Dropout-Neural-Networks/tree/master}{Enhancing Optical Diffraction Tomography with Physics Guided Adaptive Dropout Neural Networks}

\appendix

\clearpage

\begin{algorithm*}[htb]
	\DontPrintSemicolon
	\textbf{Input:}The illuminate $k_{illu}$ of $N$ LEDs, $M$ layers of imaging axial size $M\Delta z$.\\
	\textbf{Data:}Measured intensities $\{I^n_{measure}(r)\}_{n=1}^{N}$.\\
	\textbf{Hyperparameters:}The parameters of the microscope system, the step size of optimization $\alpha$, the TV regularization coefficients $\beta$. Optimization epochs: $E1$ (with Adaptive Dropout) and $E2$ (without Adaptive Dropout)\\
	\textbf{Initialization:}The RI in the layer of ADNN $\{L_m(r)\}_{m=1}^M=0$.\\
	\textbf{Return:}3D RI of the sample.\\
	\For{$i = 1:(E1+E2)$}{
		\For{$n=1:N$}{
			$k_{n} = k_{illu}^{n}$, $I_{gt}(r) = I^n_{measure}(r)$ \\
			$U_{0}(r)=\cdot exp(jk_{n}r)$ \\
			\For{$m=1:M$}{$L_m(r) \gets$ $m$th layer of ADNN \\
				$t_{m}(r,\Delta z) \gets exp(j2\pi\Delta z \dfrac{L_m(r)}{\lambda})$ \\
				$U_{m+1}(r) \gets t_{m}(r,\Delta z) \cdot \mathscr{F}^{-1}\{{P(r,\Delta z) \cdot \mathscr{F}\{U_{m}(r)\}}\}$}
		$I_{predict}(r) \gets |U_{M}(r) \mathscr{F}^{-1}\{C(u, k_0)\cdot{P(r,-\dfrac{M\Delta z}{2}) \cdot \mathscr{F}\{U_{M}\}}\}|$ \\
		$Loss \gets L_1(I_{predict}(r), I_{gt}(r))$ \\
		$Loss.autograd().backward()$ by optimizer $Adam(\alpha)$}
	$RI \gets $ layers of ADNN \\
	\If{$T<i<=E1$}{
		$RI_{reg} \gets TV_{3D}(RI,\beta)$ \\
		\For{$m=1:M$}{
			\textbf{if} $RI_{reg}(r, m\Delta z) < 0$  \textbf{dropout} $L_{m}(r)$ \\
			BPN $\gets L_{m}(r)$}}}
$RI \gets $ layers of ADNN
\caption{3D Intensity-based RI imaging with ADNN}
\end{algorithm*}

\clearpage

\begin{table}[htb]
\caption{\textbf{Hyperparameters of ADNN Optimized for Various Biological Samples}}
\begin{tabular}{ccccc}
	\toprule
	\textbf{Hyperparameters}&Neural cells&Diatom microalgae&S.spiralis&C.Elegans \\
	\midrule
	\\
	Illumination&LED matrix&LED ring&LED ring&Motorizd mirror mount \\
	\\
	Illumination units number&150&24&24&120 \\
	\\
	Learning Rate/Step Size $\alpha$&1e-3&1e-3&1e-3&1e-3\\
	\\
	Adaptive Dropout:TV Regularization $\beta$&[2,0.1,0.5]&[2,0.1,0.5]&[2,0.1,0.5]&[2,0.1,0.5]\\
	\\
	The epoch at which dropout is initiated $T$&0&5&10&0 \\
	\\
	Optimization epochs $E1$ with Adaptive Dropout &25&10&10&50 \\
	\\
	Optimization epochs $E2$ without Adaptive Dropout &25&10&10&30 \\
	\\
	\bottomrule
\end{tabular}
\end{table}

\clearpage

\begin{figure*}
\centering
\includegraphics[scale=0.18]{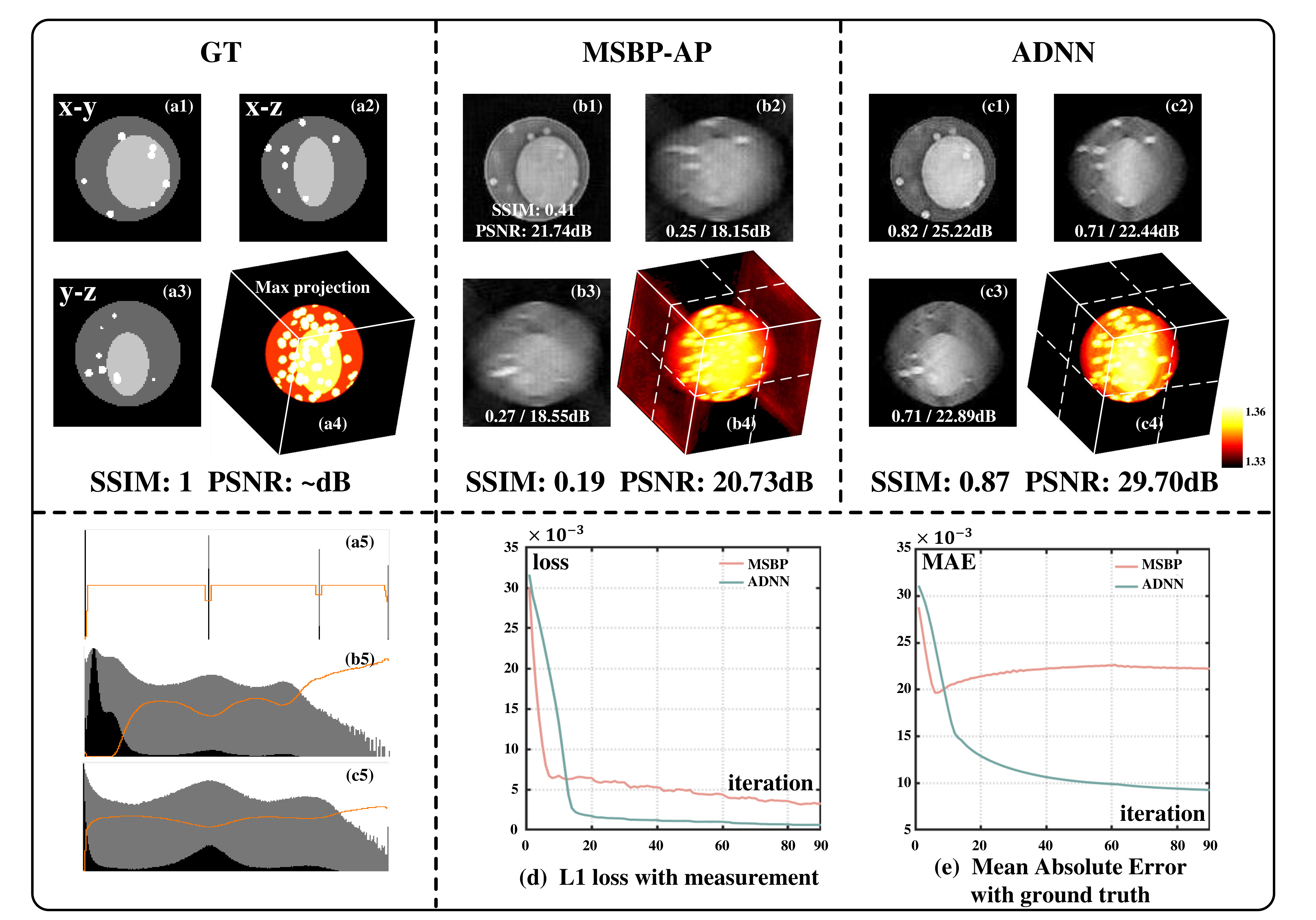}
\caption*{\textbf{Extended Data Fig.1} (a1)-(a3), (b1)-(b3), and (c1)-(c3) Reconstructed results of the central slice of the cell phantom using the ground truth, MSBP-based Alternating Projection, and ADNN methods, respectively, including SSIM and PSNR metrics compared to the ground truth. (a4), (b4), and (c4) 3D visualization results for these methods in maximum projection mode, with corresponding 3D SSIM and PSNR metrics. (a5), (b5), and (c5) Modulation functions of 3D visualizations, showing threshold modulation applied only for MSBP-AP. (d) L1 loss changes during optimization. (e) MAE during optimization.}
\end{figure*}

\clearpage

\begin{figure*}
\centering
\includegraphics[scale=0.28]{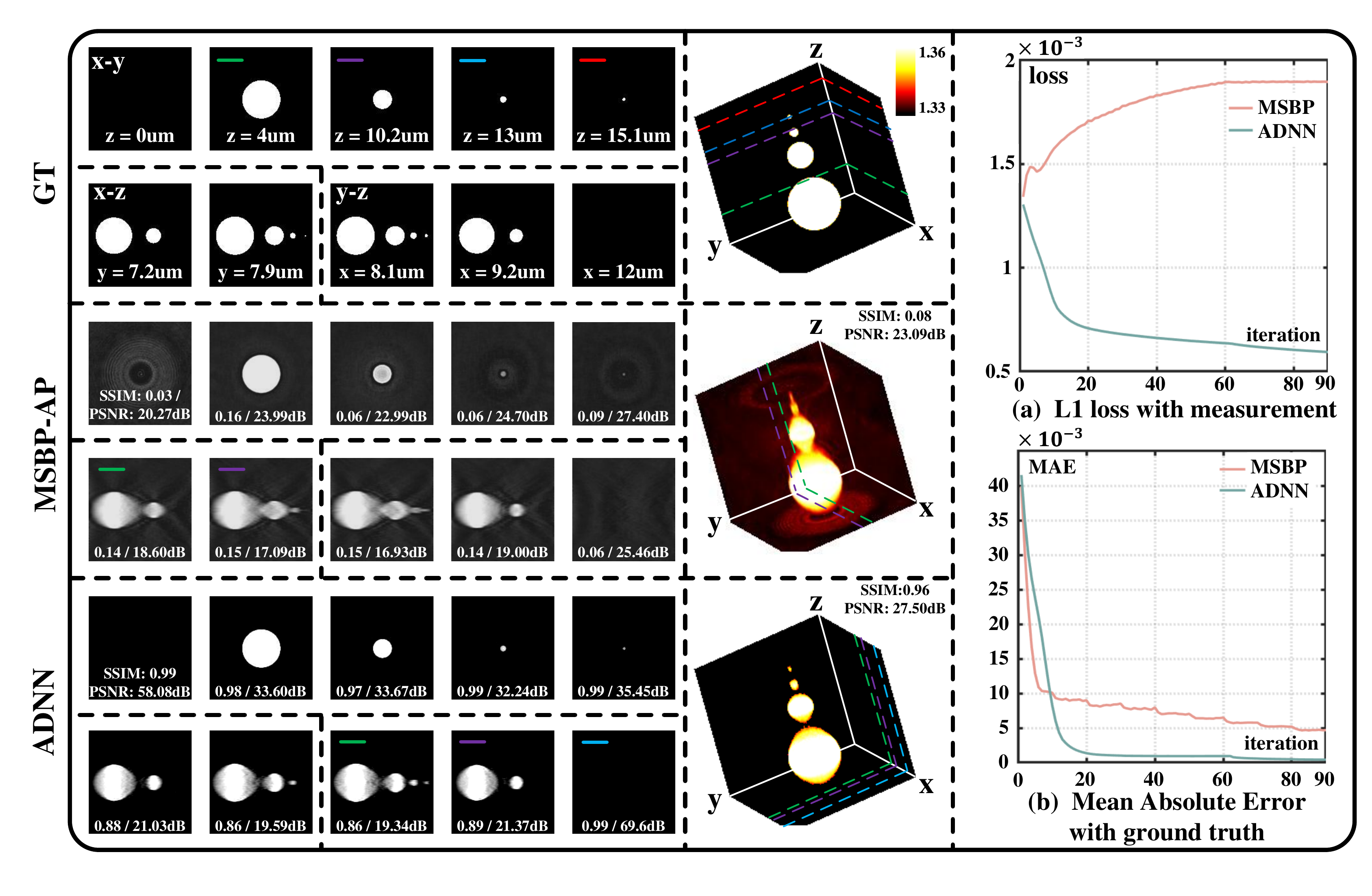}
\caption*{\textbf{Extended Data Fig.2} The simulated sample consists of microspheres arranged axially with radii of 8$\upmu$m, 4$\upmu$m, 2$\upmu$m, and 1$\upmu$m. From top to bottom, the rows correspond to the ground truth, MSBP-AP reconstructed results, and ADNN reconstructed results. (a) L1 loss changes during optimization. (b) MAE during optimization.}
\end{figure*}

\clearpage

\begin{figure*}
\centering
\includegraphics[scale=0.24]{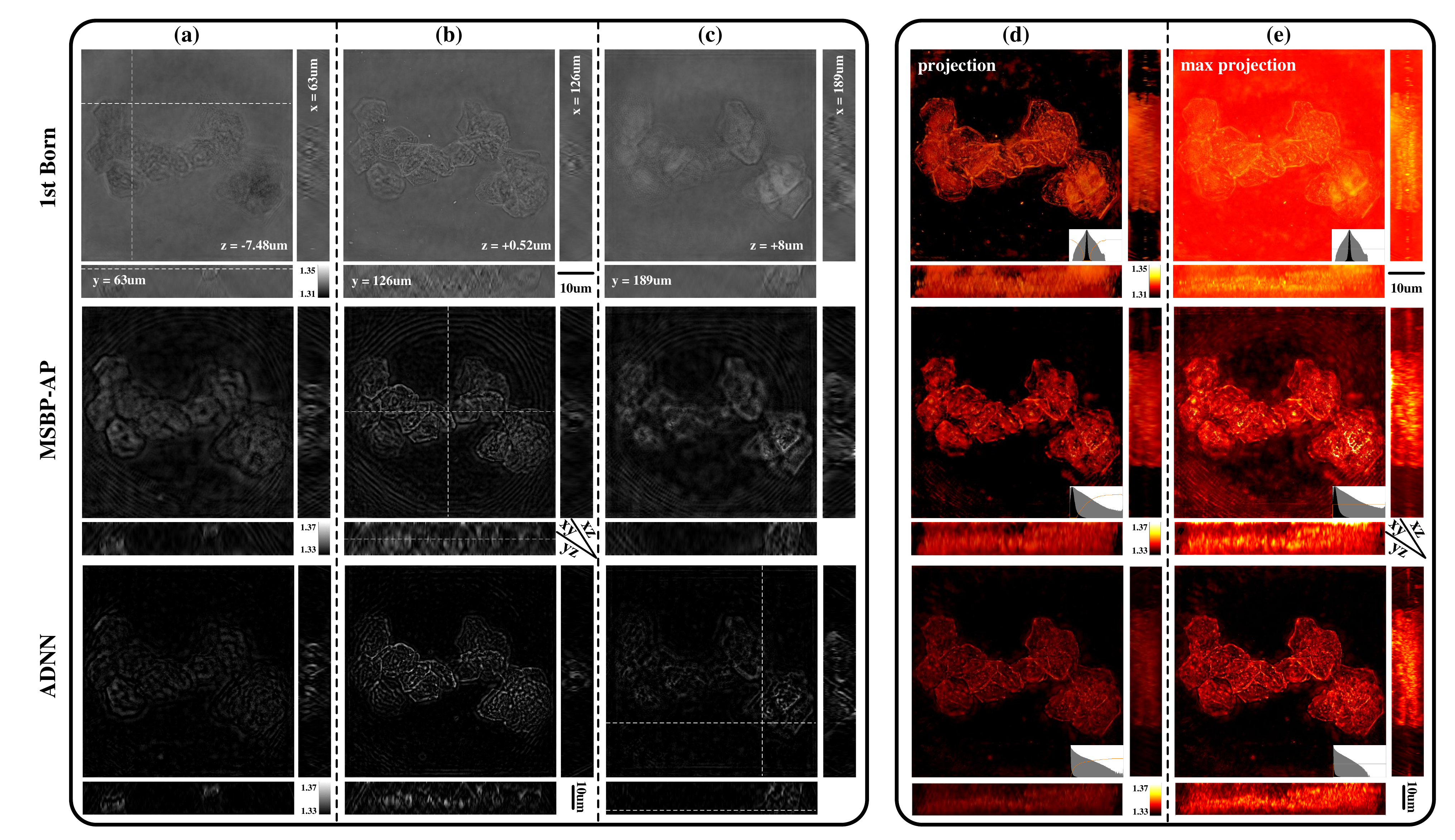}
\caption*{\textbf{Extended Data Fig.3} The tomographic reconstructed results of buccal cells. (a) Reconstructed results at z = -7.48$\upmu$m in the xy plane, x = 63$\upmu$m in the yz plane, and y = 63$\upmu$m in the xz plane for different methods. (b) Reconstructed results at z = +0.52$\upmu$m in the xy plane, x = 126$\upmu$m in the yz plane, and y = 126 $\upmu$m in the xz plane. (c) Reconstructed results at z = +8$\upmu$m in the xy plane, x = 189$\upmu$m in the yz plane, and y = 189$\upmu$m in the xz plane. (d) 3D projection results generated in ImageJ with the ID-TF curve in the bottom right corner, using the projection method. (e) 3D projection results using the max projection method.}
\end{figure*}


\clearpage

\begin{center}
	{\Large \bfseries Supplemental information: \\
		Differentiable Imaging Meets Adaptive Neural Dropout: An Advancing Method for Transparent Object Tomography} \\[1em]
\end{center}

\author{
Delong Yang \\
School of Optics and Photonics, Beijing Institute of Technology \\
Beijing, China \\
\And
Shaohui Zhang*\\
School of Optics and Photonics, Beijing Institute of Technology \\
Beijing, China \\
\texttt{*zhangshaohui@bit.edu.cn} \\
\And
Jiasong Sun \\
School of Electronic and Optical Engineering, Nanjing University of Science and Technology \\
Nanjing, China \\
\And
Chao Zuo \\
School of Electronic and Optical Engineering, Nanjing University of Science and Technology \\
Nanjing, China \\
\And
Qun Hao*\\
School of Optoelectronic Engineering, Changchun University of Science and Technology \\
Changchun, China
\texttt{*qhao@bit.edu.cn}
}


\begin{abstract} 
	This document provides supplementary information to "Differentiable Imaging Meets Adaptive Neural Dropout An Advancing Method for Transparent Object 3D Reconstruction." The document begins with a detailed explanation of the origin of the frequency domain missing cone issue in single scattering tomography methods based on the Fourier diffraction theorem. It thoroughly examines the impact that the missing cone in the frequency domain has on spatial domain reconstruction results. Subsequently, the discussion shifts to an analysis of the reasons behind the emergence of artifact issues within the multi slice beam propagation (MSBP) model based tomographic reconstruction algorithms. The text then outlines the role of dropout in the Adaptive Dropout Neural Networks (ADNN) framework and how it effectively mitigates these issues. Additionally, we present a comparative analysis of the reconstructed results from biological experiments under different display modes, followed by an explanation of the principles behind Total Variation (TV) regularization used for neuron reactivation. Finally, we elaborate on the evaluation metrics applied in the simulations, including the detailed principles of the Structural Similarity Index (SSIM) and Peak Signal-to-Noise Ratio (PSNR).
\end{abstract}

\section{Forward Scattering Model Based on 1st Born Approximation and Reconstruction Algorithm}

Based on the 1st Born approximation, the scattering model that suggests the light field through the 3D sample including the incident field and scattered field stimulated by the 3D scattering potential $V(r,z) = k^2_{0}(n^2_{media}-n^2(r,z))$, where $k_0 = 2 \pi / \lambda$. The total light field through the 3D sample can be expressed as\textsuperscript{\cite{chen2020multi,yang2023refractive}}:

\begin{gather}
	U_{in}(r, z) = \mathscr{F}^{-1}\left\{P(r,\Delta z)\cdot \mathscr{F}\{U_{in}(r, z-\Delta z)\}\right\}
	\label{Eq.1}
\end{gather}
\begin{gather}
	U^{born}_{s}(r, z) = \iiint G(r-r',\Delta z)U_{in}(r',z-\Delta z)V(r',z)d^3r'
	\label{Eq.2}
\end{gather}
The boundary condition is the incident plane wave illuminating the sample, $U_{in}(r, 0)=exp(jk_{illu}r)$ where $k_{illu}$ is the illumination wave vector at a particular angle. 
The exit electric-field, $U_{total}(r,M\Delta z)$, accounts for the light field passing through the sample. When the imaging system is focus at the center of the sample, we need to refocus the exit light field $U_{total}(r,M\Delta z)$ to the center of the sample, where $M$ denotes the total layers of the model. The final light field and intensity distributions at the image plane are:

\begin{gather}
	U_{predict}(r) = \mathscr{F}^{-1}\{C(u, k_0) ( U_{in}(r, M\Delta z) + \sum_{m}^{M} P(r,(\dfrac{M}{2}-m)\Delta z) \mathscr{F}\{U^{born}_{s}(r, m\Delta z)\})\}
	\label{Eq.3}
\end{gather}
\begin{gather}
	I_{predict}(r) = |U_{predict}(r)|^2
	\label{Eq.4}
\end{gather}
The reconstruction is to solve the following optimization with an objective consisting of a measurement loss $\mathcal{L}$ and regularizer $\mathcal{R}$:

\begin{gather}
	Argmin\{\mathcal{L}(I_{predict}(r), I_{measurement}(r)) + \mathcal{R}(r)\}
	\label{Eq.5}
\end{gather}
An alternative approach, utilizing the 1st Born approximation and Green's function method, establishes a linear relationship between the scattered field and the scattering potential of the object. This relationship in the frequency domain is known as the Fourier diffraction theorem\textsuperscript{\cite{zuo2020wide}}:

\begin{gather}
	\hat{V}(k-k{i}) = -\frac{exp(-jk_{z}z_{D})jk_{z}}{\pi}
	\label{Eq.6}
\end{gather}
\begin{gather}
	\hat{U}^{(born)}_{s}(k_{T})P(k_{T})\delta(k_{z}-\sqrt{k^2_{m}-\vert k_{T} \vert^{2}})
	\label{Eq.7}
\end{gather}
where $j$ is the imaginary unit, $k_m$ is the wave-number in the medium, and $k_i$ is the 3D wave vector of the incident plane wave. The exponential term in Eq.\ref{Eq.6} accounts for the coordinate shift in the $z$ direction and will automatically vanish if the measurement is taken at the nominal in-focus plane ($z_D = 0$). The terms $\hat{V}(k)$ and $\hat{U}^{(born)}_{s}(k_T ; z = z_D)$ represent the 3D and 2D Fourier transforms of $V(x)$ and $U^{(born)}_{s}(x_T ; z = z_D)$, respectively (where the ?hat? denotes the signal spectrum in the 2D/3D Fourier domain). The 3D frequency vector, $k = (k_T, k_z)$, lies on the 2D surface of the Ewald sphere under the constraint $k_z = \sqrt{k_m^2 - |k_T|^2}$. Consequently, the information defined by $\hat{U}^{(born)}_{s}(k_T)$ is directly related to a specific semi-spherical surface with a radius of $k_m$ in 3D Fourier space that is displaced by $-k_i$.

From the first illumination angle, select values of $\hat{f}(\mathbf{k})$ taken along its associated shell corresponding to $\mathbf{k} - \mathbf{k}_i$ and bounded by the 3D generalized aperture, $P(\mathbf{k})$ (radius of $k_0$, and maximum width of $2k_0 \text{NA}_{\text{obj}}$). The sub-region of the 3D spectrum is projected along the axial frequency coordinate to obtain a low-resolution 2D Fourier sub-spectrum $\hat{V}_i(\mathbf{k}_T)$, which is directly related to $U^{i(born)}_{s}(\mathbf{x}_T)$ after correcting for some constants according to
\begin{gather}
	\hat{U}_s^{i(born)}(\mathbf{k}_T) = -\frac{\pi j}{k_z} \hat{V}_i(\mathbf{k}_T)
	\label{Eq.8}
\end{gather}
Convert the estimated $\hat{U}^{i(born)}_{s}(\mathbf{x}_T)$ to the measured field and enforce the amplitude constraint, the update formula is:
\begin{gather}
	\hat{U}^{i(born)}_{s}(\mathbf{x}_T) \approx
	\begin{cases}
		\sqrt{I^i_c(\mathbf{x}_T)} \exp \left( U^{i(born)}_{s}(\mathbf{x}_T) + 1 \right) \left/ \left| \exp \left( U^{i(born)}_{s}(\mathbf{x}_T) + 1 \right) \right| - 1 \right. &  \\ \text{Bright-field} \\[10pt]
		\sqrt{I^i_c(\mathbf{x}_T)} \exp \left( U^{i(born)}_{s}(\mathbf{x}_T) \right) / \left| \exp \left( U^{i(born)}_{s}(\mathbf{x}_T) \right) \right| & \\ \text{Dark-field}
	\end{cases}
	\label{Eq.9}
\end{gather}

\section{Causes of Spectral Missing Cone Issues in Fourier Diffraction Methods}

In typical optical diffraction tomography (ODT) systems based on intensity measurements, data acquisition is constrained to certain angular ranges due to the inability to rotate the sample. This limitation results in missing data along the x-axis or y-axis rotations.Mathematically, this loss of data creates a missing cone in the sample's spectrum, leading to significant artifacts in the reconstructed image, particularly affecting the axial (z-axis) resolution. In practice, this manifests as elongation along the optical axis, causing objects to appear stretched or distorted, thereby degrading axial resolution.

\begin{figure}
	\centering
	\includegraphics[scale=0.27]{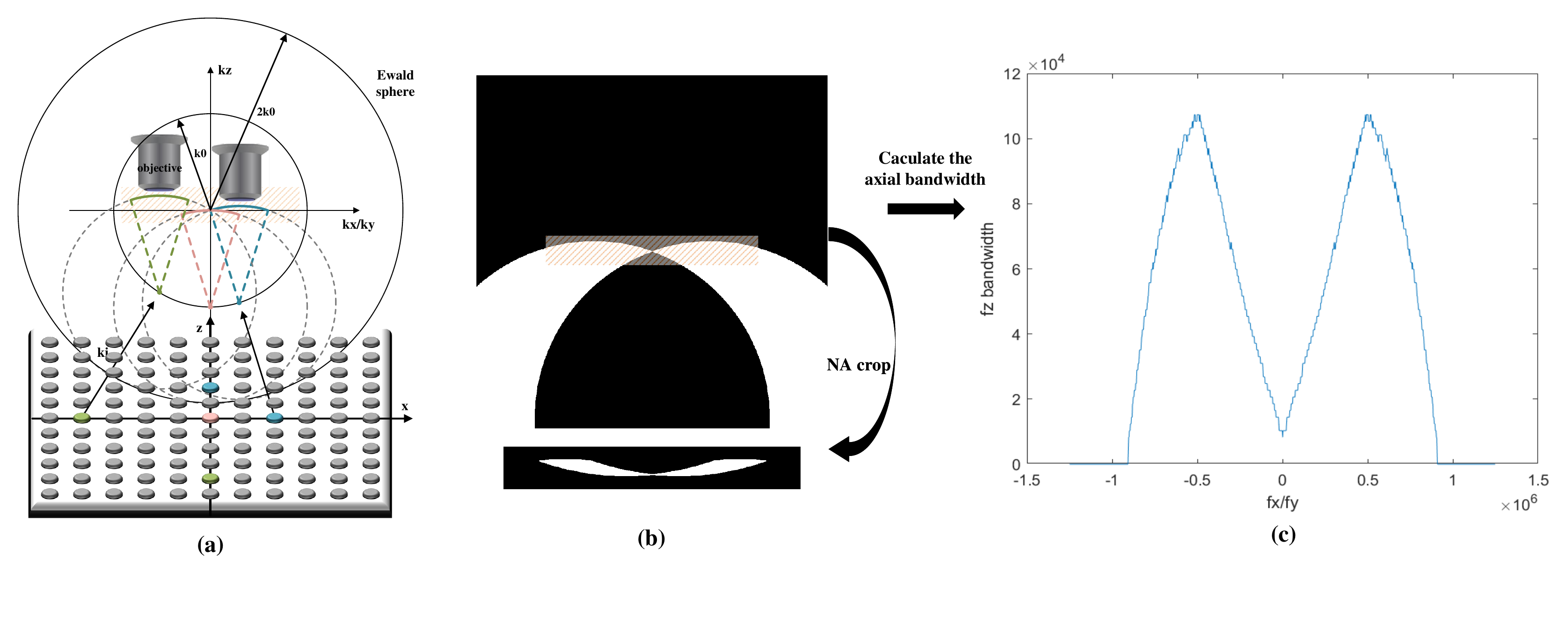}
	\caption{Fourier diffraction theorem in finite-aperture optical systems. (a) The 3D sample is illuminated by plane waves from various angles. The forward scattering wave captured by the aperture of the objective with fixed NA. (b) The bandpass on the Ewald sphere and the axial bandwidth in different lateral resolutions by the dense illumination.(c) The axial spectral bandwidth at different lateral resolutions}
	\label{Fig.1}
\end{figure}

In reconstruction methods based on the Fourier diffraction theorem, the principle of raw data collection is illustrated in Fig.\ref{Fig.1}(a). The three-dimensional(3D) spatial frequency (k-space) is defined by the 3D Fourier transform of the object's refractive index (RI)\textsuperscript{\cite{Ewald1969introduction, cowley1995diffraction}}. This spherical representation of k-space, known as the Ewald sphere, has a radius determined by the wave vector $k_{0}=\frac{2\pi}{\lambda}$. Sequential activation of each light-emitting diode (LED) at varying positions on the LED matrix provides plane waves $k_{i}$ at different angles, enabling exploration of diverse k-space regions. As $k_{i}$ varies with the illumination angle, the probed k-space is confined within a spherical shell centered on $k_{0}$. The center of this shell shifts along a second spherical shell (of radius $k_{0}$) determined by the incident angle of the plane wave $k_{i}$. However, experimental limitations imposed by the illumination and microscope optics restrict the observable k-space to a partial spherical cap, defined by the generalized aperture. Different illumination angles reposition segments of the objects k-space within the fixed numerical aperture (NA) of the microscope objective lens, allowing only a portion of the Ewald sphere to be reconstructed\textsuperscript{\cite{horstmeyer2016diffraction}}.

We employ matrices to digitally simulate the above process. From this simulation, the theoretical axial resolution is calculated by assessing the k-space bandwidth in the x-z region. The results demonstrate that different lateral resolutions correspond to varying axial resolutions. Using a microscope system equipped with a 4X objective and a 15$\times$15 LED matrix (wavelength 473 nm, bandwidth 20 nm, 100 mm from the sample) as an illustration, this system exhibits different axial bandwidths at various lateral resolutions, as shown in the \ref{Fig.1}(c) The maximum axial resolution of 10 $\upmu$m ($\Delta z = 1\times 10^5$) can be achieved for targets with a lateral size of 2$\upmu m$ ($f_x = 5\times 10^5$). For targets around 10$\upmu m$ in lateral size ($f_x = 1\times 10^5$), the axial resolution rapidly degrades to 100$\upmu m$ ($\Delta z =1\times 10^4$). This result visually demonstrates the effects caused by the missing cone problem.

\section{Layer-by-Layer Alternating Projection Reconstruction Algorithm Based on the MSBP Forward Model}
The estimate of the final intensity measurement is predicted by the forward model, $U_{predict}(\mathbf{r})$. The forward model of MSBP has already been provided in the main text. This section will explain how the conventional layer-by-layer alternating projection method (MSBP-AP) reconstructs the sample's RI information, which is mainly divided into the following ten steps\textsuperscript{\cite{chowdhury2019high}}. These ten steps are iteratively repeated until the final cost function $c(d)$ stabilizes. For notational simplicity, define $r$ and $r_{3D}$ as: $r = (x, y), r_{3D} = (x, y, z)$.

\begin{enumerate}
	\item Initialize the $N$-layer reconstruction volume with a constant RI, $n_m$. This will serve as the initial estimate of $n(r_{3D})$. Then, initialize the iteration index to $d = 0$.
	\item To start a new iteration, increment the iteration index, $d \leftarrow d + 1$, and initialize the per-iteration cost function, $c(d) = 0$.
	\item Randomly choose (without replacement) an illumination angle, with electric-field $U^i_{0}(r) = \exp(jk^i_{0} \cdot r)$ and raw intensity measurement $I^i(r)$, from the complete set of $i = 1, 2, \ldots, L$.
	\item Increment the cost function for the current iteration:
	\begin{gather}
		c(d) \leftarrow c(d) + \sum_{r} \left( \sqrt{I_{measurement}^i(\mathbf{r})} - |U_{predict}(r)| \right)^2
		\label{Eq.10}
	\end{gather}
	\item Initialize a residual term denoted by $q_{M+1}^i(r)$. For notational simplicity, define $q_0(r)$ as:
	\begin{gather}
		q_0(r) = \exp \left( j \angle U_{predict}(r) \right) \cdot \left( |U_{predict}(r)| - \sqrt{I^{measurement}(r)} \right)
		\label{Eq.11}
	\end{gather}
	\item Update $q_{M+1}^i(r)$ as:
	\begin{gather}
		q_M^i(\mathbf{r}) = \mathcal{P}_{\hat{z}} \left\{\mathscr{F}^{-1} \left\{\overline{p(k)} \cdot \mathscr{F} \left\{ q_0(\mathbf{r}) \right\} \right\} \right\}
		\label{Eq.12}
	\end{gather}
	\item For each layer of the reconstruction volume occupied by the sample, compute the back-propagation term $s_m^i(r)$ by recursively propagating backwards (i.e., $m = M, (M-1), \ldots, 2, 1$):
	\begin{gather}
		s_m^i(\mathbf{r}) = \left( -j \frac{2\pi \Delta z}{\lambda} \right) \cdot \overline{t_m(r)} \cdot \overline{\mathcal{P}_{\Delta z} \{U_{m-1}^i(r)}\} \cdot q_{m+1}^i(r)
		\label{Eq.13}
	\end{gather}
	\begin{gather}
		q_m^i(\mathbf{r}) = \mathcal{P}_{-\Delta z} \left\{\overline{t_m(r)} \cdot q_{m+1}^i(r) \right\}
		\label{Eq.14}
	\end{gather}
	\begin{gather}
		L_{m}(r) \leftarrow L_{m}(r) - \alpha \cdot s_m^i(r)
		\label{Eq.15}
	\end{gather}
	where the notation $\bar{g(r)}$ designates the complex conjugate of a 3D complex-valued variable $g(r)$. Note that Eq. (8) is the gradient-descent step for the back-propagation process, where $\alpha$ is a manually-tuned parameter to adjust the step size.
	\item Repeats steps (3)-(7) for each illumination angle $i = 1, 2, \ldots, illu num$, to incrementally refine $L_{m}(r)$ for $m = 1, 2, \ldots, M$.
	\item Implement 3D TV regularization on $L_{m}(r) (m = 1, 2, \ldots, M)$ to stabilize the iterative convergence in the presence of noise and poor conditioning, set by the parameter $\beta$:
	\begin{gather}
		n(r_{3D}) \leftarrow \text{prox}\{n(r_{3D}), \beta\}
		\label{Eq.16}
	\end{gather}
	where the operator $\text{prox}\{f(r_{3D}), \gamma\}$ is generally defined for some 3D function $f(r_{3D})$ and parameter $\gamma$ as
	\begin{gather}
		\text{prox}\{f(r_{3D}), \gamma\} = \arg \min_{g(r_{3D})} \left\{ \frac{1}{2} \|f(r_{3D}) - g(r_{3D})\|_2^2 + \gamma \text{TV}[g(r_{3D})] \right\}
		\label{Eq.17}
	\end{gather}
	where $\text{TV}[\cdot]$ denotes the 3D TV norm. The parameter $\beta$ is manually tuned to optimize the strength of the regularization. The updated $L_{m}(r)$ is the current iterative estimate of the sample's 3D RI.
	\item Repeat steps (2)-(9) to continue the iterative process until convergence is reached (when the iterative cost function $c_d$ levels out with respect to iteration index $d$).
\end{enumerate}

\section{Causes of Artifact Problem in Multi Slice Beam Propagation (MSBP) Based Reconstruction Algorithm}
In typical Optical Diffraction Tomography (ODT) systems, the MSBP-AP reconstructs spatial information directly, rather than spectral information, theoretically circumventing the traditional missing cone issue associated with frequency domain constraints. However, despite this theoretical advantage, practical implementations of MSBP-AP still exhibit axial artifacts, leading to suboptimal axial resolution. This outcome, superficially similar to the missing cone phenomenon, actually stems from the ill-posed nature of the MSBP-based inverse problem. The ill-posedness arises due to insufficient data acquisition and strong inherent correlations among the data. Current analytic optimizers, when applied to these complex, highly ill-posed inverse problems, tend to be inadequate, frequently trapping the solution in local optima. Consequently, the resulting reconstructions often display extensive axial and background artifacts, thus mirroring the characteristics of a missing cone problem, as depicted in Fig.\ref{Fig.2}.

In this paper, we detail the principles of the ADNN. To elucidate the role of dropout in ADNN, we first simplify the inverse problem to the following mathematical problem: the intensity image captured by the camera results from the accumulation along the z-axis (perpendicular to the sensor plane). During the optimization process of the inverse problem, we compute the difference between the predicted intensity distribution and the actual captured intensity distribution, and then propagate this difference back along the original direction.

We compared the conventional method, the conventional method with non-negative regularization, and the ADNN combined with dropout, as shown in Fig.\ref{Fig.2}. Both the conventional method, with or without non-negative regularization, converged only to local optima, resulting in severe axial and background artifacts. In contrast, only the ADNN integrated with dropout allowed stable convergence to the globally optimal true values.

In reality, the physical process of light propagation within the sample is more complex (detailed in the Methods section of the main text), and the actual sample is larger, corresponding to a higher number of voxels. Nonetheless, this simplified model effectively demonstrates the efficacy of our method.
\begin{figure}
	\centering
	\includegraphics[scale=0.18]{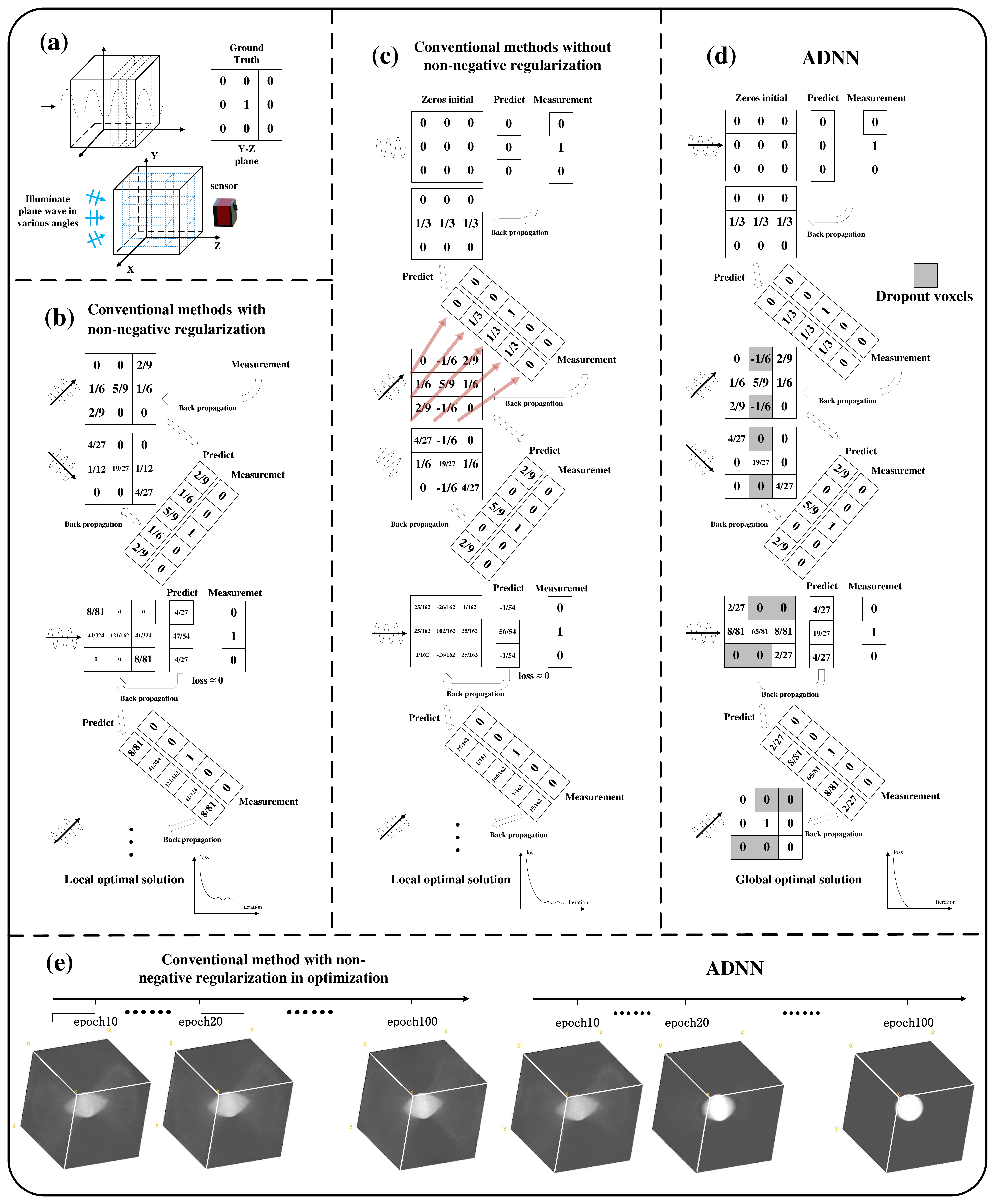}
	\caption{This figure simplifies the tomographic reconstruction process and illustrates the effectiveness of the ADNN. (a) The process of multi-angle illumination and final acquisition by the camera sensor is depicted, using an xz-plane example with a 3x3 ground truth matrix. (b) The simplified iterative process of conventional methods (MSBP-AP) with non-negative regularization. (c) Another view of the simplified iterative process of conventional methods (MSBP-AP) with non-negative regularization. (d) The simplified iterative process under the ADNN.}
	\label{Fig.2}
\end{figure}

\section{Additional 3D Projections of Reconstructed Results from Biological Experiments}
To further compare the imaging performance of different methods across the entire reconstruction space, we selected 3D projection results that had undergone subjective thresholding\textsuperscript{\cite{long2012visualization}}(based on the observer's judgment, selecting a threshold and then hiding image information below that threshold) and conducted a detailed comparison.
\begin{figure}
	\centering
	\includegraphics[scale=0.16]{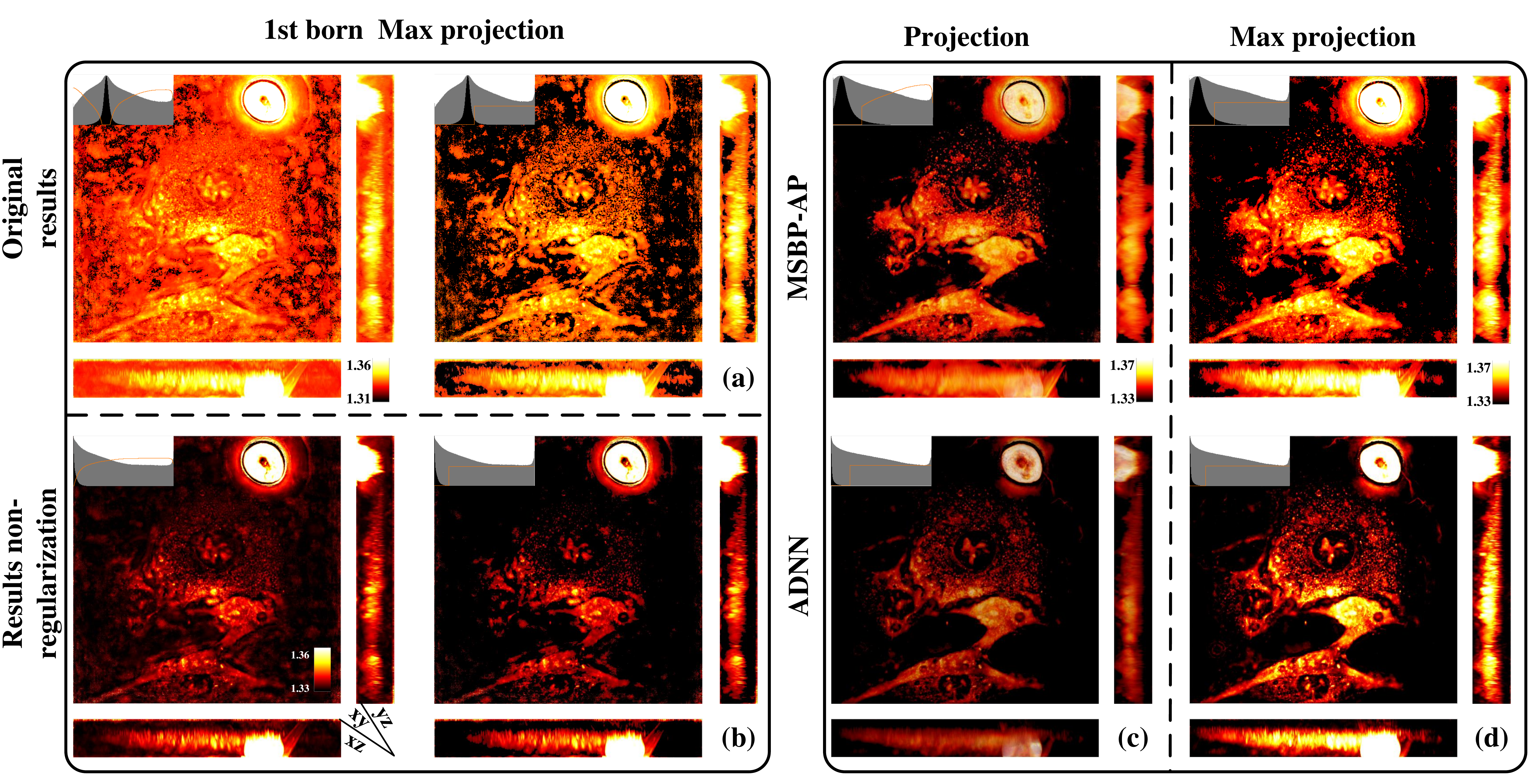}
	\caption{The 3D projection and 3D maximum projection results obtained using different reconstruction methods. The three orthogonal views (xy, yz, and xz planes) provide a comprehensive visualization of the reconstructed structures. In the top left corner of each image, the 1D-TF curve is displayed. For the projection, the 1D-TF curve represents the transparency at different refractive indices. For the maximum projection, regions where the 1D-TF curve is zero are completely transparent and hidden, while non-zero regions are fully opaque. (a) The reconstruction based on the 1st Born approximation model, with a RI range of 1.31-1.36. (b) The reconstruction based on the 1st Born approximation model, further optimized using non-negative thresholding, with a RI range of 1.33-1.36. (c) The results from the MSBP-AP, with a RI range of 1.33-1.37. (d) The reconstruction using the ADNN, with a RI range of 1.33-1.37.}
	\label{Fig.3}
\end{figure}
First, as shown in Fig.\ref{Fig.3}, the 3D projections of neuron reconstructions reveal that the ADNN exhibits minimal changes after subjective thresholding. In contrast, the 1st Born approximation method and the MSBP-based alternating projection reconstruction algorithm, although showing some improvement in signal-to-noise ratio (SNR) after subjective thresholding, suffer from significant information loss.

In addition to the three methods discussed in the main text, we also included results from the neuron experiments where non-negative thresholding was applied post-reconstruction. From an imaging perspective, this method achieves results along the axial planes (xz plane and yz plane) that are comparable to those of the ADNN. However, it is important to note that this non-negative thresholding approach may lead to inaccuracies in quantitative results. As evident in the reconstructed images, the overall RI is noticeably reduced, and severe artifacts remain around the cell nuclei and edges.
\begin{figure}
	\centering
	\includegraphics[scale=0.18]{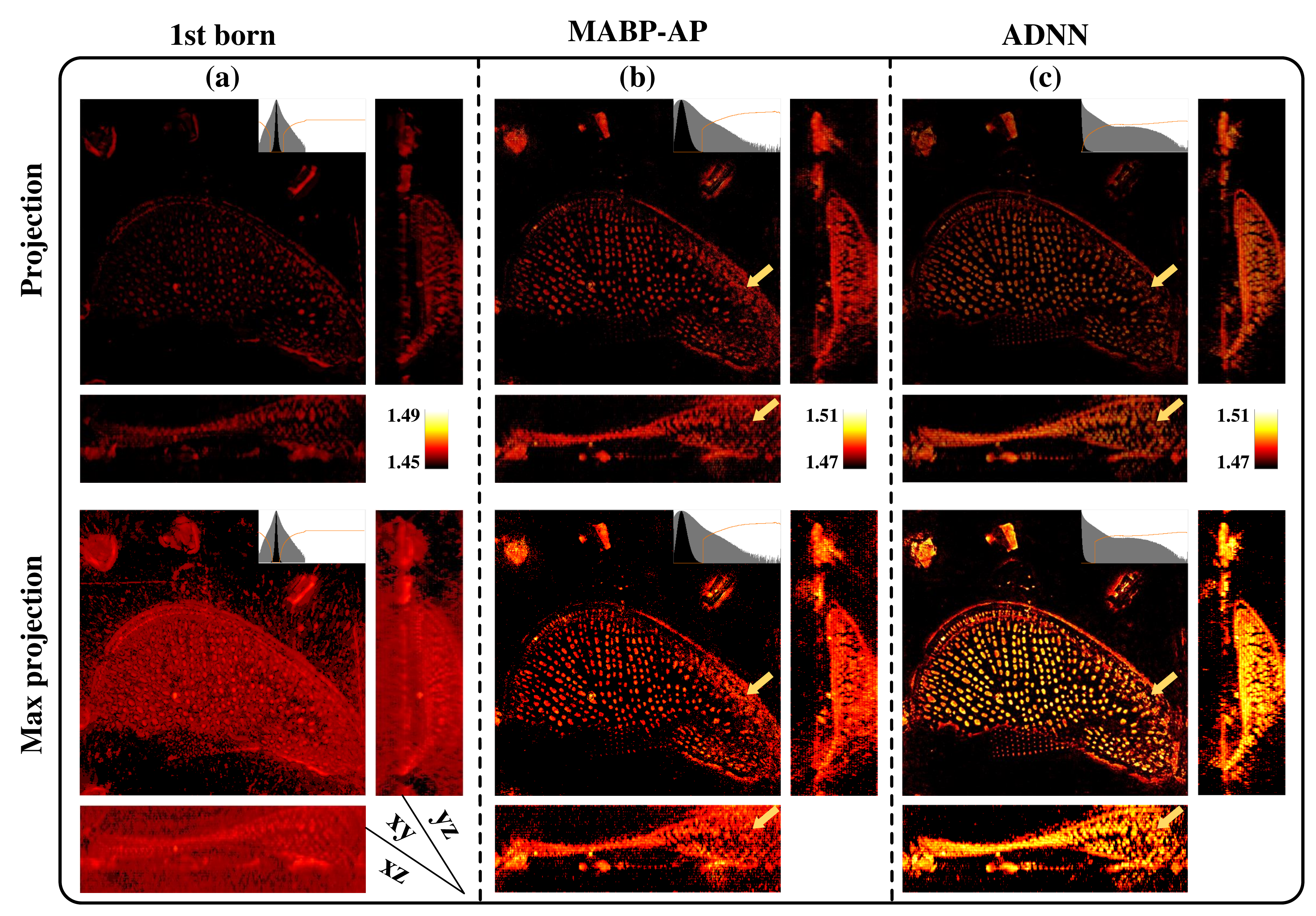}
	\caption{The 3D projection and 3D maximum projection results of the reconstructions obtained by different methods. The three orthogonal views (xy, yz, and xz planes) provide a comprehensive visualization of the reconstructed structures. In the top right corner of each image, the 1D-TF curve is displayed. (a) The reconstruction based on the 1st Born approximation model, with a RI range of 1.45-1.49. (b) The results from the MSBP-AP, with a RI range of 1.47-1.51. (c) The reconstruction using the ADNN, with a RI range of 1.47-1.51.}
	\label{Fig.4}
\end{figure}

For the diatom microalgae subjected to subjective thresholding, as shown in Fig.\ref{Fig.4}, the final results based on the 1st Born approximation still exhibit axial elongation, consistent with the findings in the main text. In regions with rapid axial position changes (as indicated by the yellow arrows), the alternating projection method based on MSBP, even after subjective thresholding, still presents excessive artifact information, making it difficult to separate these artifacts from the diatom structures. In contrast, the ADNN produces clear reconstructions in these areas.
\begin{figure}
	\centering
	\includegraphics[scale=0.27]{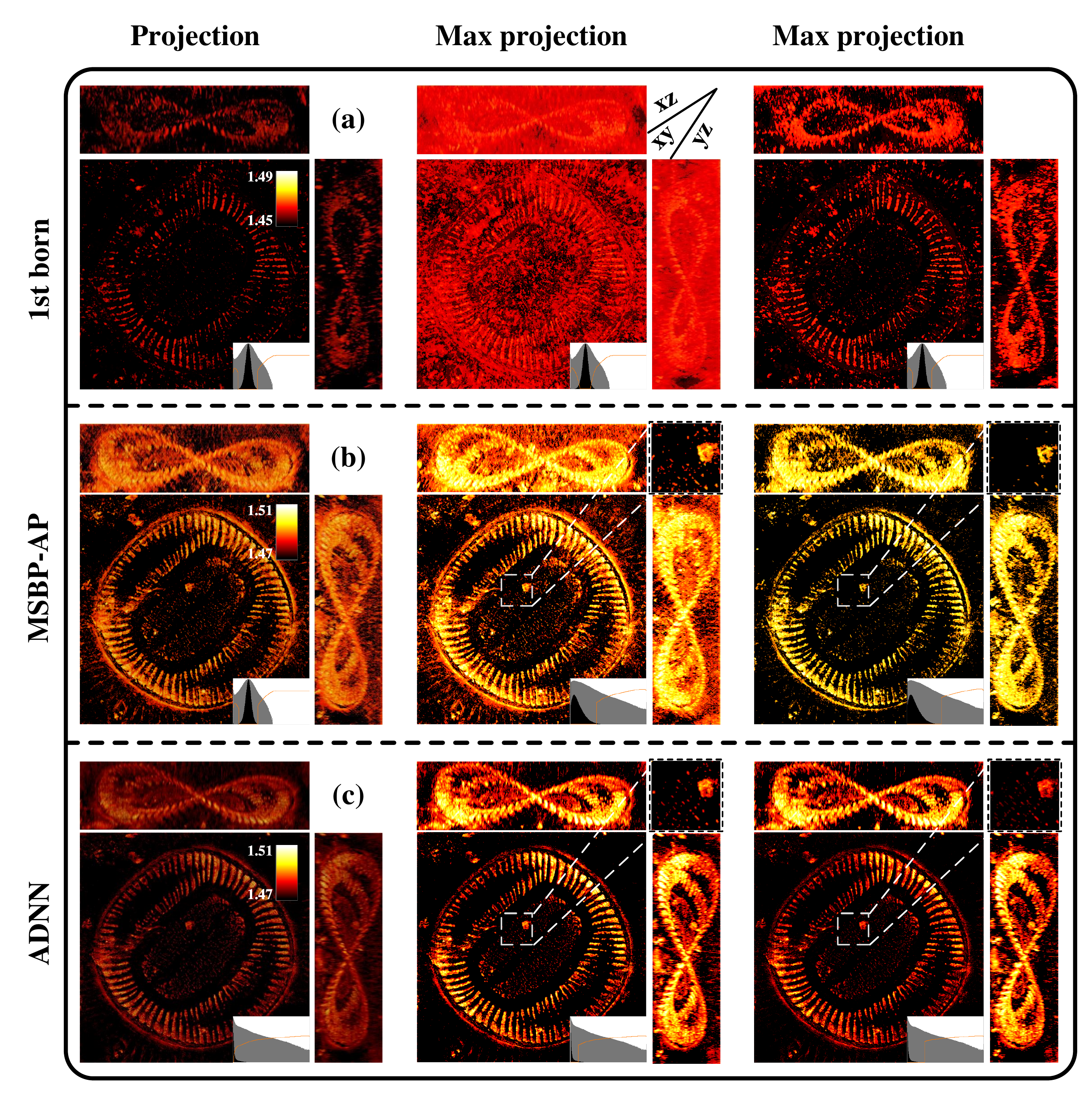}
	\caption{The 3D projection and 3D maximum projection results of the reconstructions obtained by different methods. The three orthogonal views (xy, yz, and xz planes) provide a comprehensive visualization of the reconstructed structures. In the bottom right corner of the xy plane, the 1D-TF curve is displayed. In addition to the results of 3D projection after thresholding, the 3D max projection results under two different threshold values are also shown. (a) The reconstruction based on the 1st Born approximation model, showing the 3D max projection results under two different thresholds, with a RI range of 1.45-1.49. (b) The results from the MSBP-AP, showing the 3D max projection results under two different thresholds, with a RI range of 1.47-1.51. (c) The reconstruction using the ADNN, with a RI range of 1.47-1.51.}
	\label{Fig.5}
\end{figure}

For Surirella spiralis, as shown in Figure 5, the final 3D imaging results based on the 1st Born approximation remain unsatisfactory, even after subjective thresholding. On the other hand, the MSBP-based alternating projection method, although effective in reducing noise through subjective thresholding, suffers from significant loss of high-frequency information (as highlighted in the magnified region of Figure 5), making it less effective compared to the ADNN.
\begin{figure}
	\centering
	\includegraphics[scale=0.11]{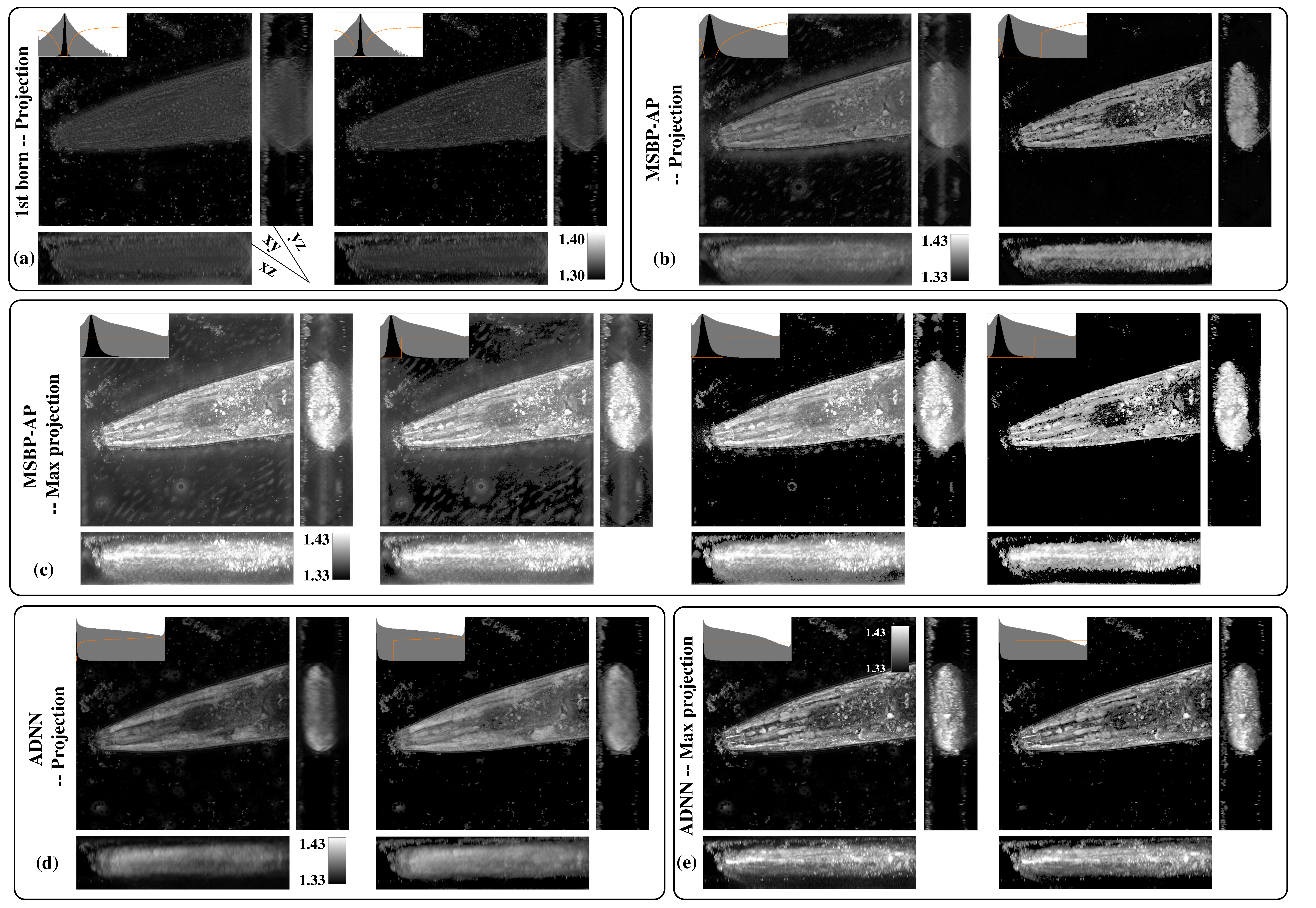}
	\caption{The 3D projection and 3D maximum projection results obtained from different reconstruction methods. The three orthogonal views (xy, yz, and xz planes) provide a comprehensive visualization of the reconstructed structures. In the top left corner of the xy plane, the 1D-TF curve is displayed. (a) The 3D projection results based on the 1st Born approximation model, with a RI range of 1.30-1.40. (b) The results from the MSBP-AP, showing the 3D projection results under two different thresholds, with a RI range of 1.33-1.43. (c) The 3D max projection results from the MSBP-AP, showing the 3D max projection results under four different thresholds, with a RI range of 1.33-1.43. (d) The 3D projection results using the ADNN, with a RI range of 1.33-1.43. (e) The 3D max projection results using the ADNN, with a RI range of 1.33-1.43.}
	\label{Fig.6}
\end{figure}

For C.elegans, which exhibits strong internal scattering, the 1st Born approximation method, after subjective thresholding, shows a significant reduction in axial and background artifacts, but the internal complex structures remain unclear. Meanwhile, the MSBP-based alternating projection method, even after multiple rounds of subjective thresholding, struggles to remove artifacts while preserving all structural details. In contrast, the ADNN, which initially produces reconstructions with minimal artifacts, shows little difference before and after subjective thresholding, with the original 3D projection already outperforming the other two methods after subjective thresholding.

\section{Total Variation (TV) Regularization Principle}
TV\cite{chambolle2004algorithm} regularization is a widely used technique in image restoration and denoising, designed to enhance image quality by suppressing noise while preserving edge information. Specifically, TV regularization introduces a regularization term that controls the strength and direction of the image update, ensuring that the restored image maintains edge sharpness while reducing unnecessary noise and details. Moreover, in our approach, TV regularization activates and preserves the information surrounding the informative voxels, preventing the erroneous discarding of relevant data during the optimization process, thereby maintaining the accuracy of low-frequency information in the reconstruction results.

In three-dimensional TV regularization, this formula is applied separately to the xy-plane, xz-plane, and yz-plane to achieve a comprehensive three-dimensional TV regularization.

The principle can be expressed by the following equation:
\begin{gather}
	I_{t+1}(i, j) = I_{t}(i, j) + \beta\left(\nabla\frac{\nabla I_{t+1}(i, j)}{|\nabla I_t(i, j)|} + \left(0.5 + \lambda\right)\left(I_0(i, j) - I_t(i, j)\right)\right)
	\label{Eq.18}
\end{gather}
In this equation: \\
\begin{itemize}
	\item $\beta$ is the regularization strength parameter, which determines the impact of the regularization term on the image update. A larger $\beta$ value enhances the regularization effect, resulting in a smoother image but potentially blurring edges, whereas a smaller  $\beta$ value reduces the regularization effect, preserving more details.
	
	\item $\nabla\frac{\nabla I_{t+1}(i, j)}{|\nabla I_t(i, j)|}$ represents the local variation characteristics of the image, controlling the smoothness and edge preservation during the process. This term plays a crucial role in noise removal while retaining the structural information of the image.
	
	\item $\lambda$ is the fidelity factor, which adjusts the degree of alignment between the original observed image $I_0(i, j)$ and the current estimated image $I_t(i, j)$ during the restoration process. By dynamically adjusting the value of $\lambda$, TV regularization can achieve a balance between noise suppression and structural preservation.
\end{itemize}

\section{Structural Similarity Index Principle}
The SSIM is a widely used metric for measuring the similarity between two images, emphasizing structural information over absolute pixel differences. SSIM is based on the idea that human visual perception is more sensitive to structural changes than to luminance or contrast alone. The SSIM index is computed by comparing three components between corresponding patches in two images: luminance, contrast, and structure. \\

1. \textbf{Luminance comparison}: The luminance \( l(x, y) \) between two image patches \( x \) and \( y \) is calculated as:
\begin{gather} 
	l(x, y) = \frac{2\mu_x \mu_y + C_1}{\mu_x^2 + \mu_y^2 + C_1} 
	\label{Eq.19}
\end{gather}
where \( \mu_x \) and \( \mu_y \) are the mean intensities of patches \( x \) and \( y \), respectively, and \( C_1 \) is a small constant to avoid division by zero. \\

2. \textbf{Contrast comparison}: The contrast \( c(x, y) \) is evaluated by comparing the standard deviations \( \sigma_x \) and \( \sigma_y \) of the two patches:
\begin{gather} 
	c(x, y) = \frac{2\sigma_x \sigma_y + C_2}{\sigma_x^2 + \sigma_y^2 + C_2} 
	\label{Eq.20}
\end{gather}
where \( C_2 \) is another small constant. \\

3. \textbf{Structure comparison}: The structure similarity \( s(x, y) \) is assessed by the correlation between the images, normalized by their standard deviations:
\begin{gather} 
	s(x, y) = \frac{\sigma_{xy} + C_3}{\sigma_x \sigma_y + C_3} 
	\label{Eq.21}
\end{gather}
where \( \sigma_{xy} \) is the covariance between patches \( x \) and \( y \), and \( C_3 \) is a constant typically set to \( C_3 = \frac{C_2}{2} \).
Finally, the SSIM index is computed as the product of these three comparisons:
\begin{gather} 
	\text{SSIM}(x, y) = \left[l(x, y)\right]^\alpha \cdot \left[c(x, y)\right]^\beta \cdot \left[s(x, y)\right]^\gamma 
	\label{Eq.22}
\end{gather}
where \( \alpha \), \( \beta \), and \( \gamma \) are parameters that control the relative importance of the three components, often set to 1. The SSIM index ranges from -1 to 1, with 1 indicating perfect structural similarity.

\section{Peak Signal-to-Noise Ratio (PSNR) Principle}
The PSNR is a commonly used metric in image processing and signal processing for evaluating the quality of a reconstructed or compressed image compared to the original. PSNR is expressed in terms of the logarithmic decibel scale and is based on the Mean Squared Error(MSE)\textsuperscript{\cite{sara2019image}} between the original and distorted images.
The Mean Squared Error between an original image \( I \) and a distorted image \( K \) of size \( m 	imes n \) is calculated as:
\begin{gather} 
	\text{MSE} = \frac{1}{mn} \sum_{i=1}^{m} \sum_{j=1}^{n} [I(i,j) - K(i,j)]^2 
	\label{Eq.23}
\end{gather}
Once the MSE is computed, the PSNR is given by the following formula\textsuperscript{\cite{hore2010image}}:
\begin{gather} 
	\text{PSNR} = 10 \cdot \log_{10} \left( \frac{L^2}{\text{MSE}} \right) 
	\label{Eq.24}
\end{gather}
where \( L \) represents the maximum possible pixel value of the image (e.g., 255 for an 8-bit image). PSNR is typically measured in decibels(dB), and a higher PSNR value indicates better image quality, as it implies that the signal has more significant variance compared to the noise.
\bibliographystyle{unsrt}
\bibliography{references}

\begin{thebibliography}{10}

\bibitem{keller2013imaging}
Philipp~J Keller.
\newblock Imaging morphogenesis: technological advances and biological
  insights.
\newblock {\em Science}, 340(6137):1234168, 2013.

\bibitem{valls2023exploring}
Arnau Valls-Esteve, N{\'u}ria Adell-G{\'o}mez, Albert Pasten, Ignasi Barber,
  Josep Munuera, and Lucas Krauel.
\newblock Exploring the potential of three-dimensional imaging, printing, and
  modeling in pediatric surgical oncology: a new era of precision surgery.
\newblock {\em Children}, 10(5):832, 2023.

\bibitem{kim2018measurements}
Geon Kim, Moosung Lee, SeongYeon Youn, EuiTae Lee, Daeheon Kwon, Jonghun Shin,
  SangYun Lee, Youn~Sil Lee, and YongKeun Park.
\newblock Measurements of three-dimensional refractive index tomography and
  membrane deformability of live erythrocytes from pelophylax nigromaculatus.
\newblock {\em Scientific reports}, 8(1):9192, 2018.

\bibitem{yla20093d}
P{\"a}ivi Yl{\"a}-Anttila, Helena Vihinen, Eija Jokitalo, and Eeva-Liisa
  Eskelinen.
\newblock 3d tomography reveals connections between the phagophore and
  endoplasmic reticulum.
\newblock {\em Autophagy}, 5(8):1180--1185, 2009.

\bibitem{agard1989fluorescence}
David~A Agard, Yasushi Hiraoka, Peter Shaw, and John~W Sedat.
\newblock Fluorescence microscopy in three dimensions.
\newblock {\em Methods in cell biology}, 30:353--377, 1989.

\bibitem{lichtman2005fluorescence}
Jeff~W Lichtman and Jos{\'e}-Angel Conchello.
\newblock Fluorescence microscopy.
\newblock {\em Nature methods}, 2(12):910--919, 2005.

\bibitem{huang2009super}
Bo~Huang, Mark Bates, and Xiaowei Zhuang.
\newblock Super-resolution fluorescence microscopy.
\newblock {\em Annual review of biochemistry}, 78(1):993--1016, 2009.

\bibitem{vicidomini2018sted}
Giuseppe Vicidomini, Paolo Bianchini, and Alberto Diaspro.
\newblock Sted super-resolved microscopy.
\newblock {\em Nature methods}, 15(3):173--182, 2018.

\bibitem{shroff2008live}
Hari Shroff, Catherine~G Galbraith, James~A Galbraith, and Eric Betzig.
\newblock Live-cell photoactivated localization microscopy of nanoscale
  adhesion dynamics.
\newblock {\em Nature methods}, 5(5):417--423, 2008.

\bibitem{huang2008three}
Bo~Huang, Wenqin Wang, Mark Bates, and Xiaowei Zhuang.
\newblock Three-dimensional super-resolution imaging by stochastic optical
  reconstruction microscopy.
\newblock {\em Science}, 319(5864):810--813, 2008.

\bibitem{hoebe2008quantitative}
RA~Hoebe, HTM Van Der~Voort, J~Stap, CJF Van~Noorden, and EMM Manders.
\newblock Quantitative determination of the reduction of phototoxicity and
  photobleaching by controlled light exposure microscopy.
\newblock {\em Journal of microscopy}, 231(1):9--20, 2008.

\bibitem{perfilov2021transient}
Maxim~M Perfilov, Alexey~S Gavrikov, Konstantin~A Lukyanov, and Alexander~S
  Mishin.
\newblock Transient fluorescence labeling: Low affinity—high benefits.
\newblock {\em International Journal of Molecular Sciences}, 22(21):11799,
  2021.

\bibitem{pawley200039}
Jim Pawley.
\newblock The 39 steps: a cautionary tale of quantitative 3-d fluorescence
  microscopy.
\newblock {\em Biotechniques}, 28(5):884--887, 2000.

\bibitem{park2018quantitative}
YongKeun Park, Christian Depeursinge, and Gabriel Popescu.
\newblock Quantitative phase imaging in biomedicine.
\newblock {\em Nature photonics}, 12(10):578--589, 2018.

\bibitem{mir2012quantitative}
Mustafa Mir, Basanta Bhaduri, Ru~Wang, Ruoyu Zhu, and Gabriel Popescu.
\newblock Quantitative phase imaging.
\newblock In {\em Progress in optics}, volume~57, pages 133--217. Elsevier,
  2012.

\bibitem{mico2008common}
Vicente Mico, Zeev Zalevsky, and Javier Garc{\'\i}a.
\newblock Common-path phase-shifting digital holographic microscopy: a way to
  quantitative phase imaging and superresolution.
\newblock {\em Optics Communications}, 281(17):4273--4281, 2008.

\bibitem{el2018quantitative}
Zahra El-Schich, Anna Leida~M{\"o}lder, and Anette Gj{\"o}rloff~Wingren.
\newblock Quantitative phase imaging for label-free analysis of cancer
  cells—focus on digital holographic microscopy.
\newblock {\em Applied Sciences}, 8(7):1027, 2018.

\bibitem{picazo2022design}
Jose~Angel Picazo-Bueno, Karina Trindade, Martin Sanz, and Vicente Mic{\'o}.
\newblock Design, calibration, and application of a robust, cost-effective, and
  high-resolution lensless holographic microscope.
\newblock {\em Sensors}, 22(2):553, 2022.

\bibitem{zhang2021review}
Jiwei Zhang, Siqing Dai, Chaojie Ma, Teli Xi, Jianglei Di, and Jianlin Zhao.
\newblock A review of common-path off-axis digital holography: towards high
  stable optical instrument manufacturing.
\newblock {\em Light: advanced manufacturing}, 2(3):333--349, 2021.

\bibitem{zheng2013wide}
Guoan Zheng, Roarke Horstmeyer, and Changhuei Yang.
\newblock Wide-field, high-resolution fourier ptychographic microscopy.
\newblock {\em Nature photonics}, 7(9):739--745, 2013.

\bibitem{jingshan2014transport}
Zhong Jingshan, Rene~A Claus, Justin Dauwels, Lei Tian, and Laura Waller.
\newblock Transport of intensity phase imaging by intensity spectrum fitting of
  exponentially spaced defocus planes.
\newblock {\em Optics express}, 22(9):10661--10674, 2014.

\bibitem{nguyen2022quantitative}
Thang~L Nguyen, Soorya Pradeep, Robert~L Judson-Torres, Jason Reed, Michael~A
  Teitell, and Thomas~A Zangle.
\newblock Quantitative phase imaging: recent advances and expanding potential
  in biomedicine.
\newblock {\em ACS nano}, 16(8):11516--11544, 2022.

\bibitem{horstmeyer2016diffraction}
Roarke Horstmeyer, Jaebum Chung, Xiaoze Ou, Guoan Zheng, and Changhuei Yang.
\newblock Diffraction tomography with fourier ptychography.
\newblock {\em Optica}, 3(8):827--835, 2016.

\bibitem{taflove2005computational}
Allen Taflove, Susan~C Hagness, and Melinda Piket-May.
\newblock Computational electromagnetics: the finite-difference time-domain
  method.
\newblock {\em The Electrical Engineering Handbook}, 3(629-670):15, 2005.

\bibitem{zhou2022transport}
Shun Zhou, Jiaji Li, Jiasong Sun, Ning Zhou, Habib Ullah, Zhidong Bai, Qian
  Chen, and Chao Zuo.
\newblock Transport-of-intensity fourier ptychographic diffraction tomography:
  defying the matched illumination condition.
\newblock {\em Optica}, 9(12):1362--1373, 2022.

\bibitem{arridge2009optical}
Simon~R Arridge and John~C Schotland.
\newblock Optical tomography: forward and inverse problems.
\newblock {\em Inverse problems}, 25(12):123010, 2009.

\bibitem{kamm2013inversion}
Jochen Kamm, Michael Becken, and Laust~B Pedersen.
\newblock Inversion of slingram electromagnetic induction data using a born
  approximation.
\newblock {\em Geophysics}, 78(4):E201--E212, 2013.

\bibitem{brown1967validity}
WP~Brown~Jr.
\newblock Validity of the rytov approximation.
\newblock {\em Journal of the Optical Society of America}, 57(12):1539--1542,
  1967.

\bibitem{wolf1969three}
Emil Wolf.
\newblock Three-dimensional structure determination of semi-transparent objects
  from holographic data.
\newblock {\em Optics communications}, 1(4):153--156, 1969.

\bibitem{gbur2002diffraction}
Greg Gbur and Emil Wolf.
\newblock Diffraction tomography without phase information.
\newblock {\em Optics letters}, 27(21):1890--1892, 2002.

\bibitem{beydoun1988first}
Wafik~B Beydoun and Albert Tarantola.
\newblock First born and rytov approximations: Modeling and inversion
  conditions in a canonical example.
\newblock {\em The Journal of the Acoustical Society of America},
  83(3):1045--1055, 1988.

\bibitem{zuo2020wide}
Chao Zuo, Jiasong Sun, Jiaji Li, Anand Asundi, and Qian Chen.
\newblock Wide-field high-resolution 3d microscopy with fourier ptychographic
  diffraction tomography.
\newblock {\em Optics and Lasers in Engineering}, 128:106003, 2020.

\bibitem{baek2021intensity}
YoonSeok Baek and YongKeun Park.
\newblock Intensity-based holographic imaging via space-domain kramers--kronig
  relations.
\newblock {\em Nature Photonics}, 15(5):354--360, 2021.

\bibitem{li2019high}
Jiaji Li, Alex Matlock, Yunzhe Li, Qian Chen, Chao Zuo, and Lei Tian.
\newblock High-speed in vitro intensity diffraction tomography.
\newblock {\em Advanced Photonics}, 1(6):066004--066004, 2019.

\bibitem{chen2020multi}
Michael Chen, David Ren, Hsiou-Yuan Liu, Shwetadwip Chowdhury, and Laura
  Waller.
\newblock Multi-layer born multiple-scattering model for 3d phase microscopy.
\newblock {\em Optica}, 7(5):394--403, 2020.

\bibitem{lim2015comparative}
JooWon Lim, KyeoReh Lee, Kyong~Hwan Jin, Seungwoo Shin, SeoEun Lee, YongKeun
  Park, and Jong~Chul Ye.
\newblock Comparative study of iterative reconstruction algorithms for missing
  cone problems in optical diffraction tomography.
\newblock {\em Optics express}, 23(13):16933--16948, 2015.

\bibitem{chowdhury2019high}
Shwetadwip Chowdhury, Michael Chen, Regina Eckert, David Ren, Fan Wu, Nicole
  Repina, and Laura Waller.
\newblock High-resolution 3d refractive index microscopy of multiple-scattering
  samples from intensity images.
\newblock {\em Optica}, 6(9):1211--1219, 2019.

\bibitem{alaei2012seismic}
Behzad Alaei.
\newblock Seismic modeling of complex geological structures.
\newblock {\em Seismic Waves-Research and Analysis}, 11:528--529, 2012.

\bibitem{kamilov2015learning}
Ulugbek~S Kamilov, Ioannis~N Papadopoulos, Morteza~H Shoreh, Alexandre Goy,
  Cedric Vonesch, Michael Unser, and Demetri Psaltis.
\newblock Learning approach to optical tomography.
\newblock {\em Optica}, 2(6):517--522, 2015.

\bibitem{matlock2023multiple}
Alex Matlock, Jiabei Zhu, and Lei Tian.
\newblock Multiple-scattering simulator-trained neural network for intensity
  diffraction tomography.
\newblock {\em Optics Express}, 31(3):4094--4107, 2023.

\bibitem{liu2022recovery}
Renhao Liu, Yu~Sun, Jiabei Zhu, Lei Tian, and Ulugbek~S Kamilov.
\newblock Recovery of continuous 3d refractive index maps from discrete
  intensity-only measurements using neural fields.
\newblock {\em Nature Machine Intelligence}, 4(9):781--791, 2022.

\bibitem{barbastathis2019use}
George Barbastathis, Aydogan Ozcan, and Guohai Situ.
\newblock On the use of deep learning for computational imaging.
\newblock {\em Optica}, 6(8):921--943, 2019.

\bibitem{yang2023refractive}
Delong Yang, Shaohui Zhang, Chuanjian Zheng, Guocheng Zhou, Yao Hu, and Qun
  Hao.
\newblock Refractive index tomography with a physics-based optical neural
  network.
\newblock {\em Biomedical Optics Express}, 14(11):5886--5903, 2023.

\bibitem{yang2022fourier}
Delong Yang, Shaohui Zhang, Chuanjian Zheng, Guocheng Zhou, Lei Cao, Yao Hu,
  and Qun Hao.
\newblock Fourier ptychography multi-parameunter neural network with composite
  physical priori optimization.
\newblock {\em Biomedical Optics Express}, 13(5):2739--2753, 2022.

\bibitem{dosovitskiy2020image}
Alexey Dosovitskiy.
\newblock An image is worth 16x16 words: Transformers for image recognition at
  scale.
\newblock {\em arXiv preprint arXiv:2010.11929}, 2020.

\bibitem{han2022survey}
Kai Han, Yunhe Wang, Hanting Chen, Xinghao Chen, Jianyuan Guo, Zhenhua Liu,
  Yehui Tang, An~Xiao, Chunjing Xu, Yixing Xu, et~al.
\newblock A survey on vision transformer.
\newblock {\em IEEE transactions on pattern analysis and machine intelligence},
  45(1):87--110, 2022.

\bibitem{hinton2012improving}
GE~Hinton.
\newblock Improving neural networks by preventing co-adaptation of feature
  detectors.
\newblock {\em arXiv preprint arXiv:1207.0580}, 2012.

\bibitem{srivastava2014dropout}
Nitish Srivastava, Geoffrey Hinton, Alex Krizhevsky, Ilya Sutskever, and Ruslan
  Salakhutdinov.
\newblock Dropout: a simple way to prevent neural networks from overfitting.
\newblock {\em The journal of machine learning research}, 15(1):1929--1958,
  2014.

\bibitem{khan2021refractive}
Rana Khan, Banat Gul, Shamim Khan, Hasan Nisar, and Iftikhar Ahmad.
\newblock Refractive index of biological tissues: Review, measurement
  techniques, and applications.
\newblock {\em Photodiagnosis and Photodynamic Therapy}, 33:102192, 2021.

\bibitem{long2012visualization}
Fuhui Long, Jianlong Zhou, and Hanchuan Peng.
\newblock Visualization and analysis of 3d microscopic images.
\newblock {\em PLoS computational biology}, 8(6):e1002519, 2012.

\bibitem{Ewald1969introduction}
PP~Ewald.
\newblock Introduction to the dynamical theory of x-ray diffraction.
\newblock {\em Acta Crystallographica Section A: Crystal Physics, Diffraction,
  Theoretical and General Crystallography}, 25(1):103--108, 1969.

\bibitem{cowley1995diffraction}
John~Maxwell Cowley.
\newblock {\em Diffraction physics}.
\newblock Elsevier, 1995.

\bibitem{chambolle2004algorithm}
Antonin Chambolle.
\newblock An algorithm for total variation minimization and applications.
\newblock {\em Journal of Mathematical imaging and vision}, 20:89--97, 2004.

\bibitem{sara2019image}
Umme Sara, Morium Akter, and Mohammad~Shorif Uddin.
\newblock Image quality assessment through fsim, ssim, mse and psnr—a
  comparative study.
\newblock {\em Journal of Computer and Communications}, 7(3):8--18, 2019.

\bibitem{hore2010image}
Alain Hore and Djemel Ziou.
\newblock Image quality metrics: Psnr vs. ssim.
\newblock In {\em 2010 20th international conference on pattern recognition},
  pages 2366--2369. IEEE, 2010.

\end{thebibliography}
\end{document}